\documentclass[a4paper,fleqn,usenatbib]{mnras}
\usepackage{newtxtext,newtxmath}
\usepackage[T1]{fontenc}
\usepackage{ae,aecompl}
\usepackage{graphicx}	
\usepackage{amsmath}	

\title[\textit{HST}/COS Lyman-alpha Absorbers in Cosmic Voids]{\textit{HST}/COS Lyman-alpha Absorbers in Cosmic Voids}

\author[W. E. Watson \& M. S. Vogeley]{
William E. Watson,$^{1}$\thanks{E-mail: wew34@drexel.edu }
and Michael S. Vogeley,$^{1}$
\\
$^{1}$Department of Physics, Drexel University, Philadelphia, PA 19104, USA\\
}

\date{Accepted XXX. Received YYY; in original form ZZZ}

\pubyear{2022}

\defcitealias{2016ApJ...817..111D}{D16}

\begin{document}
\label{firstpage}
\pagerange{\pageref{firstpage}--\pageref{lastpage}}
\maketitle

\begin{abstract}
We investigate the spatial distribution of Lyman-$\alpha$ (Ly~$\alpha$) absorbers within cosmic voids. We create a catalogue of cosmic voids in Sloan Digital Sky Survey Data Release 7 (SDSS DR7) with the VoidFinder algorithm of the Void Analysis Software Toolkit (VAST). Using the largest catalogue of low-redshift (z~$\lesssim$~0.75) IGM absorbers to date, we identify 392 Ly~$\alpha$ absorbers inside voids. The fraction of Ly~$\alpha$ absorbers inside voids (65 per cent) is comparable to the volume filling fraction of voids (68 per cent), and significantly greater than the fraction of galaxies inside voids (21 per cent). Inside voids, the spatial distribution of Ly~$\alpha$ absorbers differs markedly from that of galaxies. Galaxy density rises sharply near void edges, while Ly~$\alpha$ absorber density is relatively uniform. The radial distribution of Ly~$\alpha$ absorbers inside voids differs marginally from a random distribution. We find that lower column density Ly~$\alpha$ absorbers are more centrally concentrated inside voids than higher column density Ly~$\alpha$ absorbers. These results suggest the presence of two populations of Ly~$\alpha$ absorbers: low column density systems that are nearly uniformly distributed in the interiors of voids and systems associated with galaxies at the edges of voids.
\end{abstract}



\section{Introduction}
The intergalactic medium (IGM) has constituted the majority of baryonic matter throughout cosmic history. Though it remains challenging to detect in emission \citep{2011ApJ...736..160S, 2012MNRAS.420.1731F,2014ApJ...786..106M,2014ApJ...786..107M}, the IGM has been extensively studied through absorption-line spectroscopy of distant active galactic nuclei (AGNs), going back to the first detection by \citet{Field196}. AGN spectra display weak absorption lines blueward of the AGN frame Ly~$\alpha$ (1216 \AA) emission line. This collection of discrete absorption features, the Ly~$\alpha$ forest, is produced by diffuse intergalactic hydrogen along lines of sight to AGNs \citep{1965ApJ...142.1633G,1980ApJS...42...41S}.

The physics underlying the Ly~$\alpha$ forest has been fairly well understood since the mid 1990s, when resolved spectral studies with the Keck telescope's High Resolution Echelle Spectrometer (HIRES) matched earlier results from cold dark matter simulations \citep{1994ApJ...437L...9C,1995AJ....110.1526H}. Ly~$\alpha$ absorbers are highly photoionised clouds with temperature $\thicksim 10^4 K$ \citep{1995ApJ...453L..57Z,1996ApJ...471..582M,1996ApJ...457L..51H}. Improved understanding of intrinsic physical properties of the Ly~$\alpha$ forest has furthered understanding of the IGM. The Fluctuating Gunn-Peterson Approximation (FGPA), which relates observable Ly~$\alpha$ optical depth to intrinsic matter density \citep{1998ASPC..148...21W,1998ApJ...495...44C}, has been used to determine the matter fluctuation spectrum and mean baryon density \citep{1997ApJ...489....7R,1997ApJ...490..564W}, yielding constraints on cosmological parameters \citep{1999ApJ...520....1C, 2002ApJ...581...20C, 2000ApJ...543....1M, 2006ApJS..163...80M, 2004MNRAS.354..684V,2008PhRvL.100d1304V}.

The Ly~$\alpha$ forest has been extensively studied over an intermediate redshift range (2~$\lesssim$~z~$\lesssim$~5), where it can be observed with ground-based optical instruments. The Ly~$\alpha$ forest is more difficult to probe at lower redshift because it must be investigated with space-based UV spectrographs. Notable among the latter are Copernicus \citep{1973ApJ...181L.116S}, International Ultraviolet Explorer \citep{1978Natur.275..372B}, and Far Ultraviolet Spectroscopic Explorer \citep{2000ApJ...538L...1M}. Additionally, the Hubble Space Telescope (\textit{HST}) has hosted a series of ultraviolet spectrographs, including the Faint Object Spectrograph (FOS), Goddard High Resolution Spectrograph (GHRS), Space Telescope Imaging Spectrograph (STIS), and more recently the Cosmic Origins Spectrograph (COS). Early observations with FOS showed that the number density of Ly~$\alpha$ absorbers falls from z~$=$~2 to z~$=$~0, though not as dramatically as originally expected \citep{1991ApJ...377L...5B,1991ApJ...377L..21M}. \citet{1996ApJ...457...19B} proposed that low-z Ly~$\alpha$ absorbers might consitute a distinct population from high-z absorbers, with the former tracing galaxy halos and the latter tracing the cosmic web. Against this view, \citet{1999ApJ...511..521D} demonstrated that contemporaneous observations were entirely consistent with the structure formation model of the Ly~$\alpha$ forest. This established the prevailing consensus that low-z and high-z absorbers are fundamentally the same. 

Nevertheless, the Ly~$\alpha$ forest changes significantly with redshift. To a given sensitivity limit, the Ly~$\alpha$ forest becomes sparser with decreasing redshift. This results from the fact that a given overdensity corresponds to a higher atomic hydrogen column density N\textsubscript{HI} at high-z than at low-z \citep{1999ApJ...511..521D}. The redshift evolution of the Ly~$\alpha$ forest is dominated by expansion-driven decrease in mean baryon density and the decline of quasars, the main source of metagalactic flux \citep{1998MNRAS.297L..49T, 2001cghr.confE..64H}. By z~$\lesssim$~2 structure formation shocks have heated $\thicksim 50$ per cent of the IGM by mass above photoionisation temperature into a warm-hot (WHIM) phase \citep{1999ApJ...514....1C, 1999ApJ...511..521D}. At z~$\thicksim$~0.1, Ly~$\alpha$ absorption may be detectable in only 30~$\pm$~10 per cent of the IGM by mass \citep{2012ApJ...759...23S}. This detection shortfall lead to the long-standing ``Missing Baryon Problem'', which appears to have been recently resolved. The physical baryon density parameter $\Omega_{b} h^2$ derived from Wilkinson Microwave Anisotropy Probe (WMAP) and later Planck observations was significantly higher than direct observation showed \citep{2011ApJS..192...14J, 2016A&A...594A..22P}, until \citet{2017arXiv170910378D} and \citet{2018arXiv180504555T} located so-called "missing baryons" in a hot, diffuse phase of the IGM and the intracluster medium (ICM).

In this paper, we clarify the relationship between the low-z Ly~$\alpha$ forest and the large-scale structure seen in the spatial distribution of galaxies. As components of the IGM, Ly~$\alpha$ absorbers might be expected to trace the filaments that contain galactic groups and clusters, and to avoid the voids that occupy $\sim 70$ per cent of the volume in the universe but contain only $\sim 10$ per cent of galaxies. However, searches for voids in the high-z Ly~$\alpha$ forest have turned up negative \citep{1987MNRAS.224P..13C,1988ApJ...327L..35O, 1988A&A...197L...3P}. \citet{1998ApJ...505..506G}, using a small sample 18 Ly~$\alpha$ absorbers at very low redshift ($cz < 10^4 \ \textrm{km s}^{-1}$), found that absorbers do not trace galaxies, and appear to be distributed randomly. \citet{Penton_2002} noted that voids contain mostly low-N\textsubscript{HI} absorbers, with high-N\textsubscript{HI} absorbers tracing large-scale structure. 

More recent research at low redshift suggests that the fraction of Ly~$\alpha$ absorbers inside voids significantly exceeds the fraction of galaxies inside voids. \citet{2012MNRAS.425..245T} and \citet{panthesis} both examined \textit{HST}/STIS Ly~$\alpha$ absorbers in the SDSS DR7 galaxy void catalog of \citet{2012MNRAS.421..926P}. \citet{2012MNRAS.425..245T} stated that 43 percent of their Ly~$\alpha$ absorber sample is inside voids. They used a void catalog based on the void finding method that we use in this paper (see discussion of the VoidFinder algorithm in section \ref{sec:data} below), but to simplify their analysis, they defined each void as the largest sphere that could fit inside an irregularly shaped void region. \citet{panthesis} found 77 per cent of \textit{HST}/STIS Ly~$\alpha$ absorbers \citep{2008ApJ...679..194D} to be inside voids, and also found evidence that the absorbers prefer the deep interior volume of voids. Of the 92 absorbers inside voids, 34 were inside the inner eighth of void volume, significantly greater than the 11~$\pm$~3 that would expected for a random distribution. These striking findings motivate the research that we present below. 

We describe the galaxy void catalog and IGM absorber catalog in Section \ref{sec:data} below. In Section \ref{sec:VvsW}, we detail the classification of Ly~$\alpha$ absorbers as being in galaxy voids or walls, and compare their properties. We characterize the spatial distribution of Ly~$\alpha$ absorbers inside voids in Section \ref{sec:void_interior}. In Section \ref{sec:discussion}, we discuss our findings in light of hydrodynamic simulations that relate observables to intrinsic properties of absorbers. In Section \ref{sec:conclusion}, we summarize our results.

\section{Data}
\label{sec:data}
\subsection{Galaxies}
Our galaxy sample is derived from SDSS DR7 \citep{2009ApJS..182..543A}. A 2.5-m telescope at Apache Point Observatory in New Mexico provided photometric observations in five bands for SDSS DR7 \citep{1996AJ....111.1748F, 1998AJ....116.3040G}. Images were processed and classified, and spectroscopic observations were taken of galaxies with Petrosian r-band magnitude $r < 17.77$ \citep{1999AJ....118.1406L, 2001ASPC..238..269L, 2002AJ....124.1810S}. The galaxy catalogue we examine in this paper is Version 1.0.1 of the NASA-Sloan Atlas of \citet{2011AJ....142...31B}. 

\subsection{Ly \texorpdfstring{$\alpha$}{} Absorbers}
We use the third version of the Survey of the Low-Redshift Intergalactic Medium with \textit{HST}/COS (\citealt{2016ApJ...817..111D} hereafter D16), the largest and most sensitive survey of the low-z IGM to date\footnote{\url{https://archive.stsci.edu/prepds/igm/}}. The catalog is derived from far-UV \textit{HST}/COS spectra of 82 relatively nearby AGN (z\textsubscript{AGN}$<$0.9). The COS G130M and G160M gratings were used for 68 of the spectra, and the remaining 14 were taken with G130M alone. The spectra have high-S/N ($\gtrsim$ 15), which allows for identification of numerous weak absorption lines (down to $\log{(N_{HI})} \sim 10^{12}$). As detailed in \citetalias{2016ApJ...817..111D}, a sophisticated automated line fitting technique was used to identify Ly~$\alpha$, as well as other hydrogen, helium, and metal absorption lines in the spectra. Derived parameters include Doppler width b, redshift z\textsubscript{abs} and atomic hydrogen column density N\textsubscript{HI}. All absorbers are at redshift z\textsubscript{abs}~$<$~0.75, but Ly~$\alpha$ absorption is only observed at z~$<$~0.47; at higher redshift, Ly~$\alpha$ is redshifted past the end of the COS/G160M detector. The catalog includes both individual spectral components and redshift absorption systems, which are groupings of components within a narrow velocity window ($ 30 \ \textrm{km  s}^{-1}$ for most lines, c W\textsubscript{r} \slash $\lambda_0$ for strong broad lines) that are presumed to be physically associated. A discrete IGM cloud corresponds to an absorption system, and may display multiple spectral components. Therefore, we treat the redshift absorption systems as synonymous with Ly~$\alpha$ absorbers, and only use Ly~$\alpha$ components to compute the two point correlation function in Section \ref{sec:void_interior}.

\section{Large-scale Environment of Ly \texorpdfstring{$\alpha$}{} Absorbers}
\label{sec:VvsW}
\begin{figure}
	\includegraphics[width=\columnwidth]{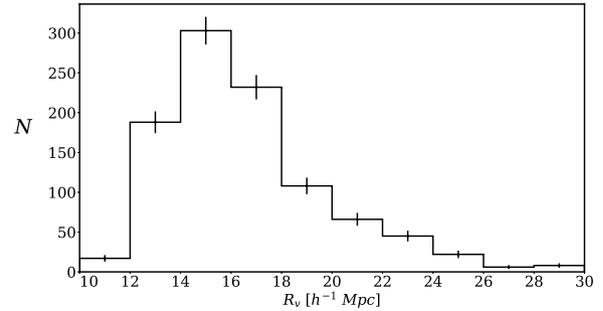}
    \caption{Histogram of effective radius for VoidFinder voids in SDSS DR7 in h\textsuperscript{-1}~Mpc. By construction, all voids have $R_{v} > 10$ h\textsuperscript{-1}~Mpc. Error bars show $\pm \sigma$ Poisson uncertainties.}
    \label{fig:reff_hist}
\end{figure}
\subsection{Galaxy Voids in SDSS DR7}
We construct a void catalog in SDSS DR7 using the VoidFinder algorithm of the Void Analysis Software Toolkit\footnote{\url{https://github.com/DESI-UR/VAST}} (VAST; \citealt{2022arXiv220201226D}). VoidFinder was originally developed by \citet{2002ApJ...566..641H, 2004ApJ...607..751H} based on the work of \citet{1997ApJ...491..421E}. The algorithm uses third nearest-neighbour distances to identify probable void galaxies, which are then removed from the sample. It then fills the cavities between the remaining galaxies with numerous overlapping spheres of various sizes. These empty spheres are combined into aspherical void regions. VoidFinder voids are constructed from the union of sets of overlapping spheres, each of which shares $\geq 50$ per cent of their volume with the largest sphere in the void. Any void too small to contain a sphere with radius 10 h\textsuperscript{-1}~Mpc is discarded. This volume cutoff means that our void catalog is unreliable within 10 h\textsuperscript{-1}~Mpc of the boundaries of SDSS DR7, and so we do not examine Ly~$\alpha$ absorbers or galaxies in that peripheral volume.
We choose the VoidFinder algorithm to define voids because comparison with tessellation-based methods clearly demonstrate that it is superior for the purpose of defining the void environment of galaxies and, presumably, of IGM absorbers (Zaidouni et al. 2022, in preparation).

We compute comoving distances of galaxies and Ly~$\alpha$ absorbers in h\textsuperscript{-1}~Mpc using the Planck 2018 estimates of $\Omega_{m}=$~0.31 and $\Omega_{\Lambda} =$~0.69 \citep{2018arXiv180706209P}. We apply VoidFinder to a volume-limited sample of SDSS DR7 galaxies (z$<$~0.107, r-band absolute magnitude M\textsubscript{r}~$<$~-20.09). Our void catalog consists of 995 galaxy voids, each of which is the union of approximately 30 non-concentric overlapping spheres of various sizes. We use Monte Carlo integration to locate the geometric center and compute the average volume of each void. For each void, we uniformly distribute random points inside a rectangular prism that contains the irregularly shaped mass of nested spheres. We eliminate the points that are outside the constituent spheres, leaving a high density sample that fills the void. The average position of that sample is the void center. Figure \ref{fig:reff_hist} is a histogram of effective void radius R\textsubscript{v}~$=$~(3V \slash 4 $\pi$ )\textsuperscript{(1/3)} for the 995 VoidFinder voids in SDSS DR7. 

We also create a sample of random points (hereafter referred to as the \textit{survey randoms}) that uniformly fills most of the three dimensional comoving volume of SDSS DR7, excluding a buffer region 10 h\textsuperscript{-1}~Mpc from the boundary surfaces of the survey, where VoidFinder cannot detect voids. We classify the survey randoms as being in walls or voids. The fraction of survey randoms inside voids, 68 per cent, approximates the volume filling fraction of VoidFinder voids in SDSS DR7. 

We note that our void catalogue is similar, but not identical, to that of \citet{2012MNRAS.421..926P}. Their void catalogue was constructed from a similar sample of SDSS DR7 galaxies to the one described in Section \ref{sec:data}. They used an earlier version of VoidFinder that had less stringent boundary conditions than the VAST VoidFinder that we use for our void catalogue (see \citealt{2022arXiv220201226D} for details). In the \citet{2012MNRAS.421..926P} void catalogue, constituent spheres of voids may extend outside the galaxy survey by as much as 50 per cent of their volume. By contrast, the constituent spheres of voids in our catalogue cannot have more than 10 per cent of their volume outside the SDSS DR7 volume. Note again that our statistical analyses do not use any objects that lie within a 10 h\textsuperscript{-1}~Mpc buffer near the survey boundary.

The developers of the VAST package have recently published a new void catalog for SDSS DR7 using the newest version of VoidFinder \citep{2022arXiv220201226D}. They used VoidFinder on the same sample of SDSS DR7 galaxies that we use, but with a less restrictive magnitude and redshift cut on the volume-limited sample (M\textsubscript{r}$<$-20, z$<$0.114). As a result, their catalog has more voids and spans a somewhat larger volume within SDSS DR7. 

\subsection{AGN Sight Lines in SDSS DR7}
\label{sec:agnsightlines}

\begin{figure}
	\caption{Sky projection in equatorial coordinates of the SDSS DR7 footprint (orange), with positions of the 36 \textit{HST}/COS AGN sight lines that pierce it (crosses). Name, abbreviated name (as denoted in \citetalias{2016ApJ...817..111D}), right ascension $\alpha$ and declination $\delta$ in degrees, and AGN redshift z\textsubscript{AGN} of sight lines are listed in the table below.}
	\includegraphics[width=1.\columnwidth]{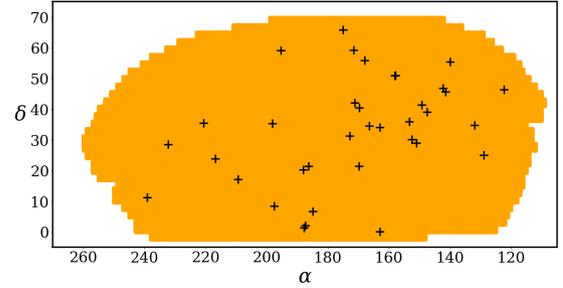}
	\begin{tabular}{ | l | l | l | l | l |}
		Name & Abbr. & z\textsubscript{AGN} & $\alpha$ & $\delta$\\
		\hline
    		  SDSS J080908.13+461925  &  s080908  &  0.66  &  122.28  &  46.32 \\
    		  PG 0832+251  &  pg0832  &  0.33  &  128.9  &  24.99 \\
    		  PG 0844+349  &  pg0844  &  0.06  &  131.93  &  34.75 \\ 
    		  Mrk 106  &  mrk106  &  0.12  &  139.98  &  55.36 \\ 
    		  SDSS J092554.43+453544  &  s09255b  &  0.33  &  141.48  &  45.6 \\ 
    		  SDSS J092909.79+464424  &  s092909  &  0.24  &  142.29  &  46.74 \\ 
    		  SDSS J094952.91+390203  &  s094952  &  0.37  &  147.47  &  39.03 \\ 
    		  PG 0953+414  &  pg0953  &  0.23  &  149.22  &  41.26 \\ 
    		  PG 1001+291  &  pg1001  &  0.33  &  151.01  &  28.93 \\ 
    		  FBQS J1010+3003  &  f1010  &  0.26  &  152.5  &  30.06 \\ 
    		  Ton 1187  &  ton1187  &  0.08  &  153.26  &  35.86 \\ 
    		  1ES 1028+511  &  1es1028  &  0.36  &  157.83  &  50.89 \\ 
    		  1SAX J1032.3+5051  &  1sj1032  &  0.17  &  158.07  &  50.86 \\ 
    		  PG 1048+342  &  pg1048  &  0.17  &  162.93  &  33.99 \\ 
    		  PG 1049-005  &  pg1049  &  0.36  &  162.96  &  0.0 \\ 
    		  HS 1102+3441  &  hs1102  &  0.51  &  166.42  &  34.43 \\ 
    		  SBS 1108+560  &  sbs1108  &  0.77  &  167.88  &  55.79 \\ 
    		  PG 1115+407  &  pg1115  &  0.15  &  169.63  &  40.43 \\ 
    		  PG 1116+215  &  pg1116  &  0.18  &  169.79  &  21.32 \\ 
    		  PG 1121+422  &  pg1121  &  0.22  &  171.16  &  42.03 \\ 
    		  SBS 1122+594  &  sbs1122  &  0.85  &  171.47  &  59.17 \\ 
    		  Ton 580  &  ton580  &  0.29  &  172.79  &  31.23 \\ 
    		  3C 263  &  3c263  &  0.65  &  174.99  &  65.8 \\ 
    		  PG 1216+069  &  pg1216  &  0.33  &  184.84  &  6.64 \\ 
    		  PG 1222+216  &  pg1222  &  0.43  &  186.23  &  21.38 \\ 
    		  3C 273  &  3c273  &  0.16  &  187.28  &  2.05 \\ 
    		  Q 1230+0115  &  q1230  &  0.12  &  187.71  &  1.26 \\ 
    		  PG 1229+204  &  pg1229  &  0.06  &  188.02  &  20.16 \\ 
    		  PG 1259+593  &  pg1259  &  0.48  &  195.3  &  59.04 \\ 
    		  PG 1307+085  &  pg1307  &  0.16  &  197.45  &  8.33 \\ 
    		  PG 1309+355  &  pg1309  &  0.18  &  198.07  &  35.26 \\ 
    		  SDSS J135712.61+170444  &  s135712  &  0.15  &  209.3  &  17.08 \\ 
    		  PG 1424+240  &  pg1424  &  0.6  &  216.75  &  23.8 \\ 
    		  Mrk 478  &  mrk478  &  0.08  &  220.53  &  35.44 \\ 
    		  Ton 236  &  ton236  &  0.45  &  232.17  &  28.42 \\ 
    		  1ES 1553+113  &  1es1553  &  0.41  &  238.93  &  11.19 \\
		\hline
	\end{tabular}
    \label{fig:sky}
\end{figure}

We restrict our investigation to AGN sight lines within the SDSS DR7 footprint, which includes 36 of the 82 AGN sight lines examined by \citetalias{2016ApJ...817..111D} (see Figure \ref{fig:sky}). The mean effective radius R\textsubscript{v} of VoidFinder voids in SDSS DR7 is 16~h\textsuperscript{-1}~Mpc, roughly 20 times smaller than the average comoving length of the sight lines inside SDSS DR7. Thus, we expect each sight line to pierce multiple void and wall regions, and we can consider the angular positions of the AGN sight lines to be unbiased with respect to large-scale structures. However, we have a relatively small number of sight lines along which we can probe the IGM, so it is possible that, by chance, these sight lines do not probe a representative sample of large-scale structures. To characterize how these 36 sight lines intersect large-scale structure in SDSS DR7, we construct dense a linear random distribution along the lines of sight and we determine if each random point lies inside a VoidFinder void. We refer to this distribution as the \textit{sight line randoms}. 

\subsection{Ly \texorpdfstring{$\alpha$}{} Absorbers and Galaxies in SDSS DR7 Voids}
The sight lines in Figure \ref{fig:sky} contain 605 absorption systems inside the redshift limit of our void catalog (0.003~$\leq$~z~$\leq$~0.107). If an absorber lies inside any of the constituent spherical holes of the voids, it is classified as a void absorber. Otherwise, it is considered a wall absorber. Galaxies are classified in the same manner. Since VoidFinder voids have a minimum radius of 10~h\textsuperscript{-1}~Mpc, some of the objects we classify as being in walls actually reside in small voids. In our classification scheme, walls encompass a wide range of environmental conditions, from dense structures to small voids. The void classification is stricter.

\begin{figure*}
	\includegraphics[width=2.\columnwidth]{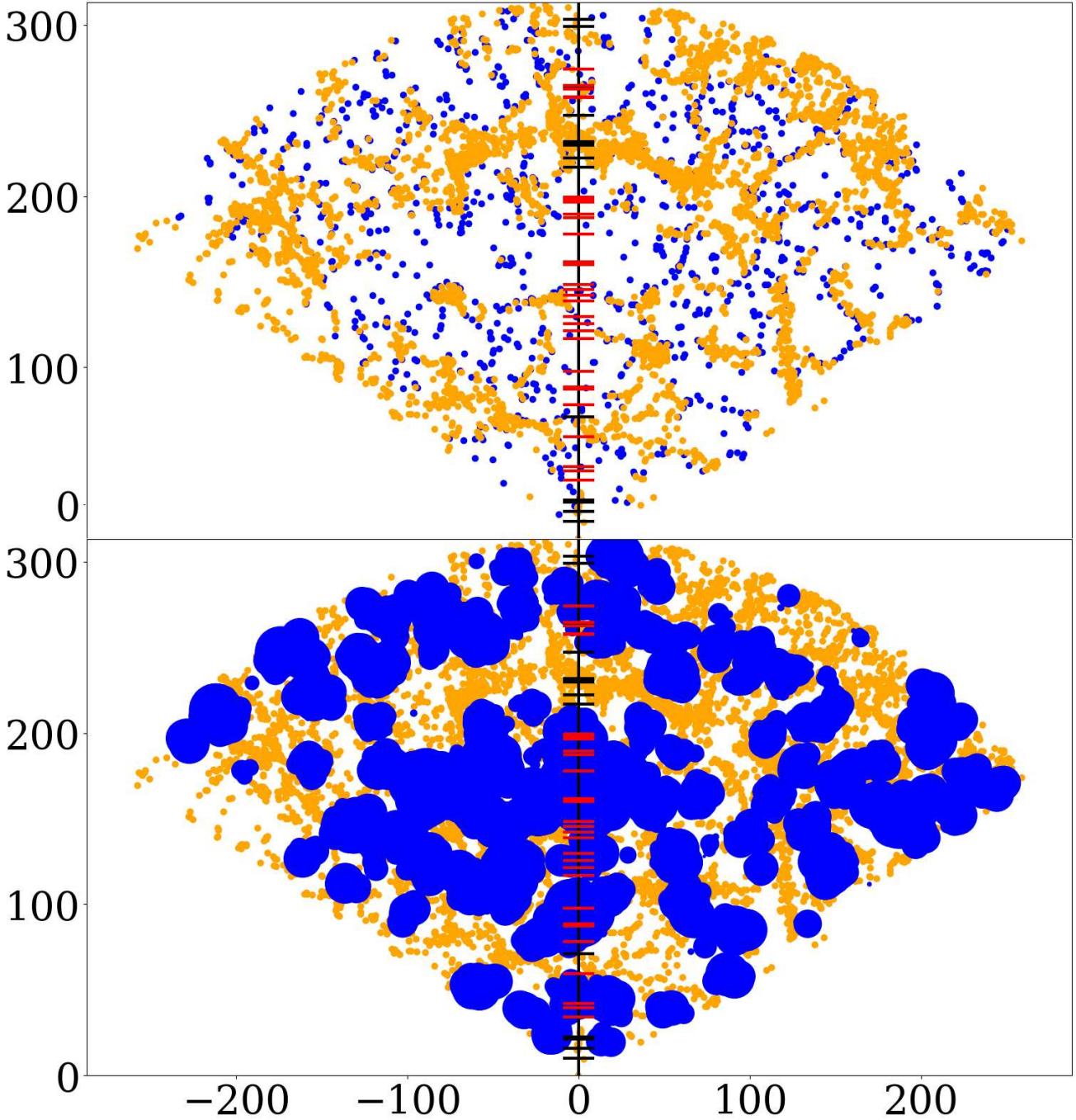}
    \caption{\textit{Top} - A 10~h\textsuperscript{-1}~Mpc thick slice of SDSS DR7 in comoving Cartesian coordinates. The vertical axis is aligned with the line of sight to AGN Q1230+0115, indicated by the vertical line, and the slice midplane contains the sight line. Wall and void galaxies are marked respectively in orange and blue, while wall and void Ly~$\alpha$ absorbers are black and red bars along the sight line. The AGN itself is outside the redshift limit of SDSS DR7. \textit{Bottom} - The same slice from SDSS DR7, showing the intersection of VoidFinder voids with the slice midplane in blue. Axes in h\textsuperscript{-1}~Mpc.}
    \label{fig:slice}
\end{figure*}

We illustrate results of our classification in Figure \ref{fig:slice} (a), a 10~h\textsuperscript{-1}~Mpc thick slice of SDSS DR7, with wall and void galaxies from the volume-limited sample in orange and blue, respectively. The slice midplane contains the sight line to AGN Q1230+0115, indicated by a black line, with void and wall Ly~$\alpha$ absorbers marked as red and black bars, respectively. Figure \ref{fig:slice} (b) shows the same slice, with irregularly shaped blue blobs representing VoidFinder voids. Each VoidFinder void is the union of a set of spheres (about 30 per void).  Figure \ref{fig:slice} (b) plots in grey the overlapping circular intersections of the constituent holes with the slice midplane.

We find that Ly~$\alpha$ absorbers are not preferentially found in walls, unlike galaxies, which suggests that the spatial distribution of Ly~$\alpha$ absorbers does not trace filaments and clusters in the galaxy distribution. The fraction of Ly~$\alpha$ absorbers in voids is 65 per cent, comparable to the fraction of sight line randoms in voids, 72 per cent, which approximates the fractional length of the AGN sight lines inside voids. In contrast, only 21 per cent of SDSS DR7 galaxies in the volume-limited sample are inside the voids. Unsurprisingly, the galaxy void fraction is far lower than the fraction of survey randoms inside voids, which approximates the filling fraction of voids, 68 per cent. The small difference in void fraction of sight line randoms and survey randoms, 72 vs. 68 per cent, reflects the finite sampling effect discussed above in Section \ref{sec:agnsightlines}.

\section{Spatial Distribution of Ly \texorpdfstring{$\alpha$}{} Absorbers in Voids}

\label{sec:void_interior}
\begin{figure*}
\centering
    \includegraphics[width=1.\textwidth]{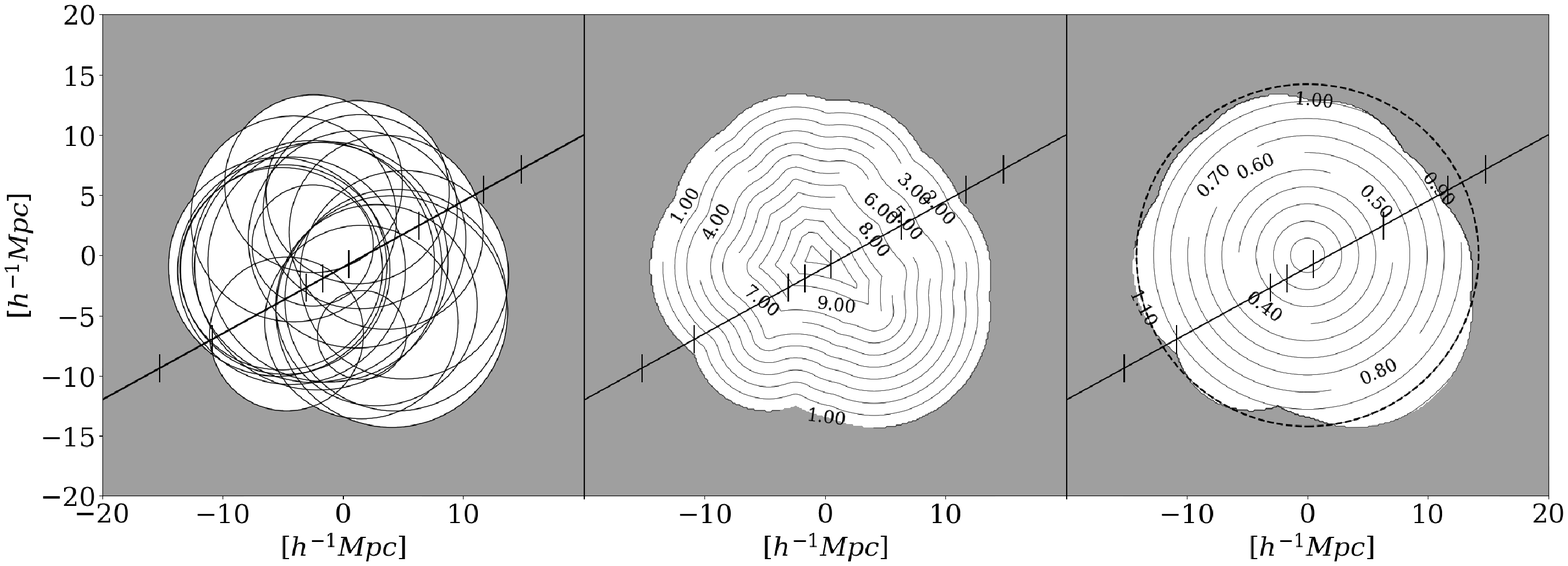}
    \caption{Three views of the same thin slice of a void in SDSS DR7. Axes show comoving coordinates. A simulated AGN sight line pierces the void, with Ly~$\alpha$ absorber positions marked by vertical black ticks. The origin of this plot sits on the void's volume-weighted centre. \textit{Left -} Constituent spherical holes of a VoidFinder void, which have circular intersections with the slice. This void is typical of others in our catalogue in that it consists of about 30 overlapping holes. The void is the union of the holes, and its intersection with the slice is shown in white. We emphasize that this void, like all VoidFinder voids that consist of multiple holes, does not have a perfectly spherical boundary. \textit{Centre -} Contours of distance to the walls d\textsubscript{w} in h\textsuperscript{-1}~Mpc. We use d\textsubscript{w} to characterize the positions of void absorbers and galaxies in the stacked density profiles shown in Figure \ref{fig:rho_dwall}. This is useful for understanding environmental effects of proximity to walls, as d\textsubscript{w} is sensitive to the irregular void boundary. \textit{Right -} Circular annuli of scaled voidcentric distance, r\slash R\textsubscript{v}, emanating from the void centre. The dashed circle indicates the effective radius R\textsubscript{v}. We use r\slash R\textsubscript{v} to characterize the positions of void absorbers and galaxies in the stacked density profiles shown in Figure \ref{fig:rho_rnorm}. This helps us to understand environmental conditions of the interior of a void, but is less useful at higher values of r\slash R\textsubscript{v}, where the annuli start to collide with the irregular void boundary.    
}
    \label{fig:voidslice}
\end{figure*}

Theoretical considerations of how physical conditions of the IGM vary with large-scale structure suggest that we carefully examine how Ly~$\alpha$ absorbers are spatially distributed.
Cosmological hydrodynamic simulations indicate that the Warm-Hot Intergalactic Medium (WHIM) becomes a more important constituent of the IGM at z<1 \citep{2010MNRAS.408.2051D}. We expect that the WHIM is predominantly found in and around walls, because the IGM in voids, particularly in the deep interiors of voids, does not experience gravitational shock-heating from structure formation. Additionally, super-Hubble expansion inside voids should adiabatically cool and lower the density of the IGM. Thus, Hydrogen inside voids should be cooler, less ionised, and more diffuse than in walls. These large-scale environmental effects on the physical state of intergalactic hydrogen affect the spatial distribution of Ly~$\alpha$ absorbers. Here we examine how the density of Ly~$\alpha$ absorbers varies inside voids, from the boundaries with the walls down into the centers of voids. 

To examine enivornmental effects on IGM, we need to carefully characterize void shape and see where absorbers reside. The left panel of Figure \ref{fig:voidslice} shows the intersection of a void and its constituent spherical holes with a plane. This void, like others in our catalogue, has a complex, irregular geometry. The centre and right panels compare two measures of position inside the void, the distance from the walls d\textsubscript{w} and the scaled voidcentric distance r\slash R\textsubscript{v}. At low d\textsubscript{w} (see Figure \ref{fig:rho_dwall}), effects from shock-heating in walls should be more prominent. Near the void centre, it is perhaps more informative and straightforward to consider concentric shells of r\slash R\textsubscript{v}. We measure the distance outward from the volume-weighted centre of the void, and divide that distance by the effective void radius represented by the dashed black circle. Shells of r\slash R\textsubscript{v} are spherical near the centre, but start to get cut off as they collide with the void boundary at higher r\slash R\textsubscript{v}. We keep track of which objects are inside the void, so the absorber at upper right, which is just outside the void but less than R\textsubscript{v} from the centre, would not be considered when we construct the density profiles shown in Figure \ref{fig:rho_rnorm}. We note that previous research on the void density profile (e.g. \citealt{2004ogci.conf...51C}) has used simple spherical models for void shape. This slice shows that such an approach tends to mix wall and void regions together at the void boundary.

\subsection{Wall Distance}
The IGM at the edges of voids, where they interface with dense structures, may be shock-heated by structure formation, albeit less strongly than in the walls themselves. Thus, the IGM near void boundaries is likely to be more highly ionised than the IGM deeper inside voids. At the same time, the void density profile of the IGM should behave similarly to that of the volume-limited sample of SDSS DR7 galaxies shown in Figure \ref{fig:rho_dwall}, with a sharp increase around a voidcentric distance of R\textsubscript{v}. 
We compute the density of Ly~$\alpha$ absorbers as a function of distance d\textsubscript{w} from cosmic walls to observe the effect of variation in IGM density and ionisation fraction on Ly~$\alpha$ absorbers.

\begin{figure}
	\includegraphics[width=\columnwidth]{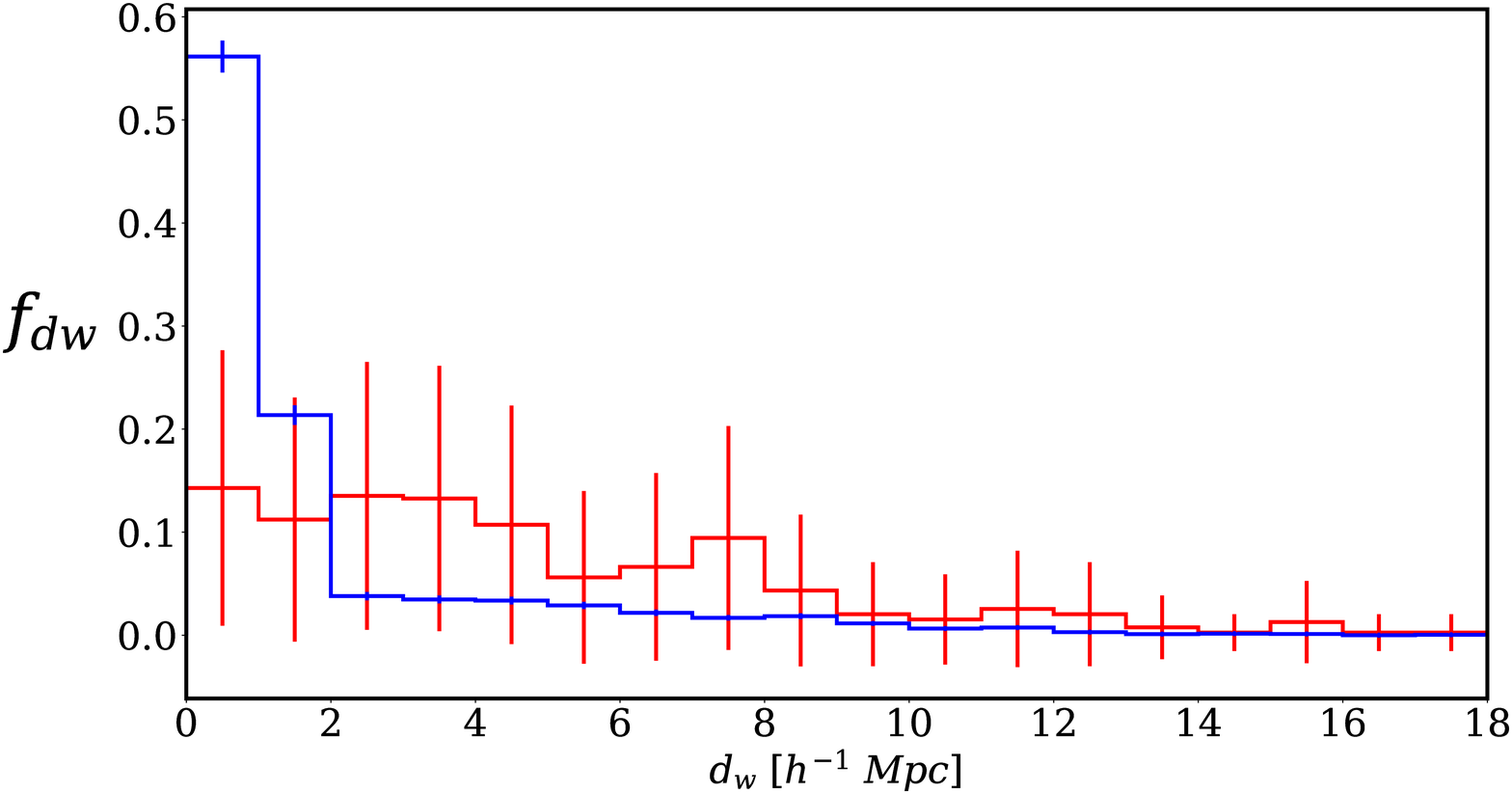}
    \caption{Normalized histograms of distance to cosmic wall $d_w$ for Ly~$\alpha$ absorbers (red) and galaxies (blue) inside voids. There are about 50 times as many void galaxies as void Ly~$\alpha$ absorbers. Error bars show $\pm \sigma$ Poisson uncertainties.}
    \label{fig:N_dwall}
\end{figure}

\begin{figure}
	\includegraphics[width=\columnwidth]{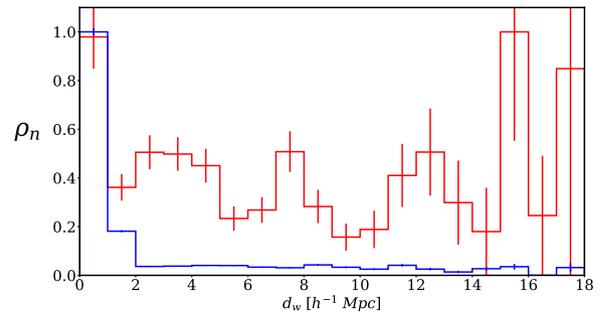}
    \caption{Density profile of Ly$\alpha$ absorbers (red) and galaxies (blue) inside voids as a function of distance to cosmic wall in h\textsuperscript{-1}~Mpc. The normalized density $\rho_n$ on the vertical axis is a unitless number proportional to the number density along the sight lines. Profiles are stacked for all voids, which range in effective radius from 10 to 30~h\textsuperscript{-1}~Mpc.}
    \label{fig:rho_dwall}
\end{figure}

To compute the distance from cosmic walls d\textsubscript{w} to each void Ly~$\alpha$ absorber, we generate a uniform distribution of random points on the irregular boundary of each void (recall that the volume of each void is the union of overlapping spheres of varying radius). We estimate d\textsubscript{w} as the distance from a void Ly~$\alpha$ absorber to the nearest random point on the void surface. In the same way, we compute d\textsubscript{w} for void galaxies. Histograms of d\textsubscript{w} for void Ly~$\alpha$ absorbers and void galaxies shown in Figure \ref{fig:N_dwall}, are markedly different. Void galaxies are almost all within 2 h\textsuperscript{-1}~Mpc of cosmic walls. The number of void Ly~$\alpha$ absorbers shows a shallower decline with d\textsubscript{w}. 

To convert the measurements of d\textsubscript{w} to estimates of densities of Ly~$\alpha$ absorbers and galaxies inside voids, we also compute d\textsubscript{w} for survey randoms and sight line randoms (see Section \ref{sec:VvsW}) inside voids. This allows us to estimate the available void volume as a function of d\textsubscript{w}, which we use to determine the number density of galaxies as a function of d\textsubscript{w} for each void. We then stack the galaxy density profiles as shown in Figure \ref{fig:rho_dwall} (blue). We also show the stacked density profile of Ly~$\alpha$ absorbers (red). We estimate length of AGN sight lines inside bins of d\textsubscript{w} with the sight line randoms, and use that to construct a linear density profile of Ly~$\alpha$ for the voids. In these void density profiles, the bar height is normalized by the maximum density in voids, so that the vertical axis $\rho_n$ is unitless, but still proportional to the physical density.

The linear density of Ly~$\alpha$ absorbers in Figure \ref{fig:rho_dwall} (red) spikes near the void boundary, but otherwise shows no clear trend with d\textsubscript{w}. The density peaks at 15-16~h\textsuperscript{-1}~Mpc and 17-18~h\textsuperscript{-1}~Mpc are statistically insignificant, as these bins each contain fewer than five absorbers. The volume density of void galaxies is greatest near void-wall interfaces, but falls precipitously beyond d\textsubscript{w}~$=$~1~h\textsuperscript{-1}~Mpc and remains relatively flat thereafter.

\subsection{Voidcentric Distance}
Next we examine how the densities of Ly~$\alpha$ absorbers and galaxies vary from the centers of voids outwards. We measure these density profiles both as a function of physical distance in h\textsuperscript{-1}~Mpc and as a function of the scaled distance from each void center. To estimate densities, we use randoms to estimate the available volume in each bin of distance, as we did above for the density dependence on d\textsubscript{w}. This is necessary because, as noted above, we include only objects inside voids and volume inside voids to compute density profiles.

\begin{figure}
	\includegraphics[width=\columnwidth]{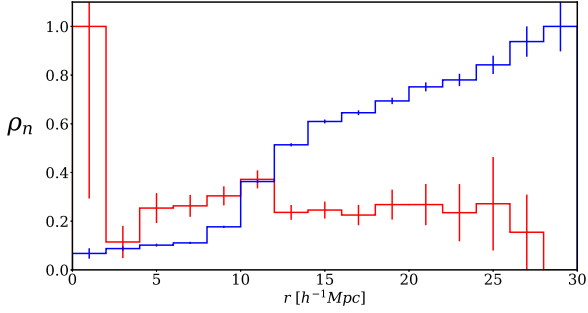}
    \caption{Radial density profile of Ly~$\alpha$ absorbers (red) and galaxies (blue) inside cosmic voids as a function of voidcentric distance, r, in h\textsuperscript{-1}~Mpc. The normalized density $\rho_n$ on the vertical axis is a unitless number proportional to the physical density. Profiles are stacked for all voids, which range in effective radius from 10 to 30 h\textsuperscript{-1}~Mpc. The density of Ly~$\alpha$ is relatively flat (the enhancement in the lowest bin is not statistically significant). However, the rough statistical test summarized in Table \ref{tab:stattest} below shows that absorbers in the lowest quartile of column density are over-represented in the deep interiors of voids, while the highest quartile of column density is under-represented. By contrast, the density of galaxies is $\lesssim$ 10 per cent of the cosmic mean near void centres and rises sharply near void edges.}
    \label{fig:rho_r}
\end{figure}

Figure \ref{fig:rho_r} shows the void density profiles of Ly~$\alpha$ absorbers and galaxies as functions of voidcentric distance r. The profiles are stacked for all voids, and the bar height is normalized to eliminate physical units as before. Note that a particular value of voidcentric distance, r, does not correspond to a fixed value of wall distance, $d_{w}$, because voids are not spherically symmetric. As discussed in Section \ref{sec:data}, the geometric center of each void is located using Monte Carlo integration. The density of galaxies increases with voidcentric distance, with an abrupt rise between 10 and 12 h\textsuperscript{-1}~Mpc. The Ly~$\alpha$ absorber density profile, by contrast, is relatively flat; the prominent spike in the lowest bin of r is not statistically significant, because it encompasses a very small volume and contains only two Ly~$\alpha$ absorbers.

\begin{figure}
	\includegraphics[width=\columnwidth]{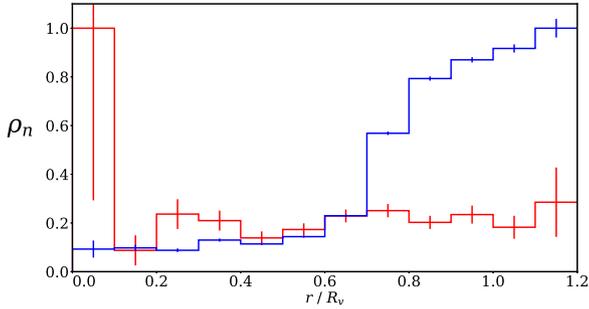}
    \caption{Radial density profile of Ly~$\alpha$ absorbers (red) and galaxies (blue) inside cosmic voids as a function of scaled voidcentric distance, r\slash R\textsubscript{v}, where R\textsubscript{v} is the effective void radius. The normalized density $\rho_n$ on the vertical axis is a unitless number proportional to the physical density. Profiles are stacked for all voids, which range in effective radius from 10 to 30~h\textsuperscript{-1}~Mpc. The Ly~$\alpha$ density is relatively flat. The peak near the void centre, around r\slash R\textsubscript{v}~$=$~0, is not statistically significant because only 2 absorbers fall in this bin. This behavior is significantly different from the radial density profile of void galaxies, which is lowest in the interior volume of voids ($r/R_{v} <$ 1/2) and rises near the void edges.}
    \label{fig:rho_rnorm}
\end{figure}

To compare density profiles of voids with different sizes, we compute the scaled voidcentric distance r\slash R\textsubscript{v} for each Ly~$\alpha$ absorber and galaxy inside voids. In Figure \ref{fig:rho_rnorm}, we show the void density profile of Ly~$\alpha$ absorbers (red) as a function of r\slash R\textsubscript{v}. Density is generally relatively flat. The  enhanced Ly~$\alpha$ absorber density near the void center is of low statistical significance. This behavior is quite distinct from that of void galaxies (blue). Galaxy density is lowest near the void centers and rises dramatically near void-wall boundaries. From Figure \ref{fig:rho_dwall}, we might expect an even sharper rise of galaxy density near the void boundaries, r\slash R\textsubscript{v}~$\approx$~1, but that rise is softened, because voids are not spherical.

\subsubsection{Statistics of Void Ly \texorpdfstring{$\alpha$}{} and Galaxy Spatial Distributions}
\begin{figure}
	\includegraphics[width=\columnwidth]{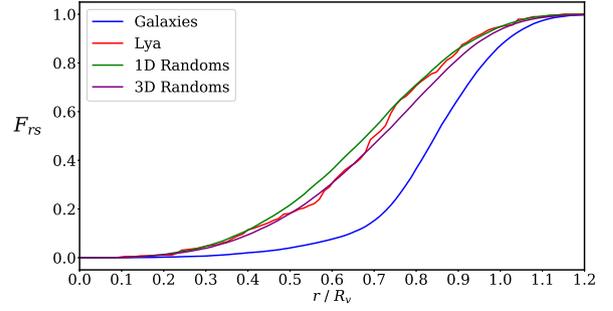}
    \caption{Cumulative distributions of Ly~$\alpha$ absorbers (red), galaxies (blue), and a Poisson distribution (green) as functions of scaled voidcentric distance r\slash R\textsubscript{v}.}
    \label{fig:cdf}
\end{figure}

To quantify the statistical significance of differences between the galaxy and Ly~$\alpha$ absorber distributions inside voids, we compute their cumulative distributions $F_{rs}$ as functions of scaled voidcentric distance r\slash R\textsubscript{v}, as shown in Figure \ref{fig:cdf}. We also compute the cumulative distribution functions of the sight line randoms (see Section \ref{sec:VvsW}), as well as a random distribution of uniform volume density. Figure \ref{fig:cdf} shows that  Ly~$\alpha$ absorbers (red) are very slightly less centrally concentrated in voids than linear randoms (green), but significantly more centrally concentrated than galaxies (blue). 

We perform an Anderson-Darling test on the cumulative distribution functionss of Ly~$\alpha$ absorbers and the sight line randoms. We find a p-value of approximately six per cent, a marginally significant result that suggests that Ly~$\alpha$ absorbers are not uniformly distributed inside voids. This quantifies the visual impression in Figure~\ref{fig:cdf} that Ly~$\alpha$ absorbers are slightly less centrally concentrated inside voids than sight line randoms. This difference in cumulative distributionss is most apparent at scaled voidcentric distances of $0.4$ to $0.7$. 

We also carry out the Anderson-Darling test to compare galaxies with Ly~$\alpha$ absorbers, and find a statistically significant difference (p$<$0.01); galaxies are more concentrated near the void-wall interface around r\slash R\textsubscript{v}~$=$~1, whereas the Ly~$\alpha$ absorbers have a more uniform distribution inside voids. This statistical result is further evidence that galaxies and Ly~$\alpha$ absorbers have distinct spatial distributions. 

\subsubsection{Two-Point Correlation Function of Ly \texorpdfstring{$\alpha$}{} Absorbers}
\begin{figure}
	\includegraphics[width=\columnwidth]{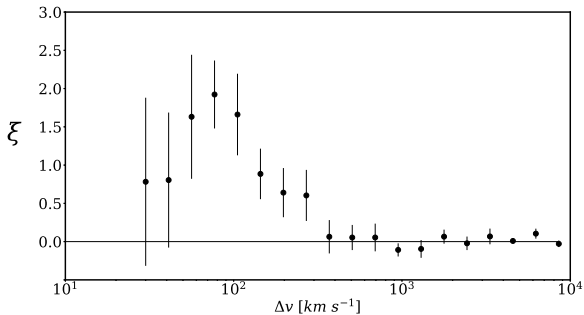}
    \caption{The two-point correlation function $\xi$ of Ly~$\alpha$ absorption lines in the \textit{HST}/COS low-z IGM survey. The lines are only weakly clustered even over small velocity separations, so the positions of lines can be regarded as statistically independent.}
    \label{fig:tpcf}
\end{figure}

We compute the two-point correlation function (2PCF) of Ly~$\alpha$ lines to characterize their small-scale clustering. This statistic was presented in \citetalias{2016ApJ...817..111D} using an earlier version of the \textit{HST}/COS low-z IGM catalog. Here, we use a similar procedure to find the 2PCF using the latest edition of the catalog. Individual Ly~$\alpha$ absorption lines were used to compute the 2PCF, since redshift systems group closely spaced components together and thus provide no information about small-scale clustering. The estimator of the 2PCF is,
\begin{equation}
\xi (\Delta v) = \frac{N_{obs}(\Delta v)}{N_{ran}(\Delta v)} - 1
\end{equation}
where N\textsubscript{obs} and N\textsubscript{ran} are the number of pairs of observed and simulated random absorbers within a velocity separation $\delta v$ \citep{2010arXiv1009.1232L}. Thus $\xi$ gives the probability of clustering above random. The sight line randoms that we use to compute N\textsubscript{ran} are distributed uniformly along the AGN sight lines. By contrast, \citetalias{2016ApJ...817..111D} employ a more complicated procedure to distribute random absorbers. At each resolution element $\Delta v$ in an AGN spectrum, they compute the minimum equivalent width for detection based on the S/N of the data to determine the probability of detection within the resolution element. They place a random absorber within the resolution element if the probability exceeds a random number. Notwithstanding this difference in the randoms, our resulting 2PCF for Ly~$\alpha$ absorbers, shown in \ref{fig:tpcf}, is nearly identical to that presented by \citetalias{2016ApJ...817..111D}. Clustering is weak, especially for $\Delta v > 100\ {\rm km \ s}^{-1}$, which indicates that there is not significant large-scale structure in the spatial distribution of Ly~$\alpha$ absorbers. By contrast, galaxies typically exhibit strong clustering as measured by the 2PCF in redshift space, $\xi \approx $~5 on scales r~$<$~10~h\textsuperscript{-1}~Mpc \citep{2006MNRAS.368...21L}. 

\subsection{Doppler Parameters of Void and Wall Ly \texorpdfstring{$\alpha$}{} Absorbers}
Simulations \citep{2011ApJ...741...99C} indicate that structure formation shock heats the intergalactic medium in walls but not voids. This environmental effect could be reflected in the Doppler \textit{b} parameter, a measure of the width of an absorption line, which is a function of not only absorber temperature, but also nonthermal effects. In some cases, the COS spectral resolution may not be fine enough to resolve component structure, particularly for strong absorption systems.  Following the example of \citetalias{2016ApJ...817..111D}, we examine the Doppler parameter of weak (log(N\textsubscript{HI}) $<$ 13.5), single-component systems for which the b-value is more indicative of temperature. The mean b-value is somewhat higher in voids (mean $\bar{b} =$ 32.4  $\pm$ 1.2 km s\textsuperscript{-1}, median 28.0 km s\textsuperscript{-1}) than walls (mean $\bar{b}=30.6 \pm$ 1.8 km s\textsuperscript{-1}, median 25.2 km s\textsuperscript{-1}). This result is surprising, given that we might expect Ly~$\alpha$ absorbers in the walls to have a higher average b-parameter because of shock heating from structure formation. However, the difference is of modest statistical significance and the resolution of HST/COS is predicted to limit the use of the $b$ parameter for measuring thermal properties of the IGM in this range of line width \citep{2010MNRAS.408.2051D}.

\subsection{Column Density and Spatial Distribution of Void Ly \texorpdfstring{$\alpha$}{} Absorbers}
We test for dependence on the large-scale void environment of the observed column density, N\textsubscript{HI}, of absorption systems. This distribution is of interest because, while the void environment is characterized by lower matter density and, therefore, structures with smaller characteristic mass, the IGM in voids is expected to have a higher neutral fraction. The statistical distribution of column densities reflects a complex, environmentally-sensitive interplay of astrophysics.

Perhaps surprisingly, comparison of the means of the column density $N(HI)$ of absorbers observed in walls and voids does not reveal a strong environmental difference.
We find that the mean $\log{N_{HI}}$ of wall absorbers, 13.59~$\pm$~0.89, does not significantly differ from that of void absorbers, 13.43~$\pm$~0.73. 

\begin{figure}
	\includegraphics[width=\columnwidth]{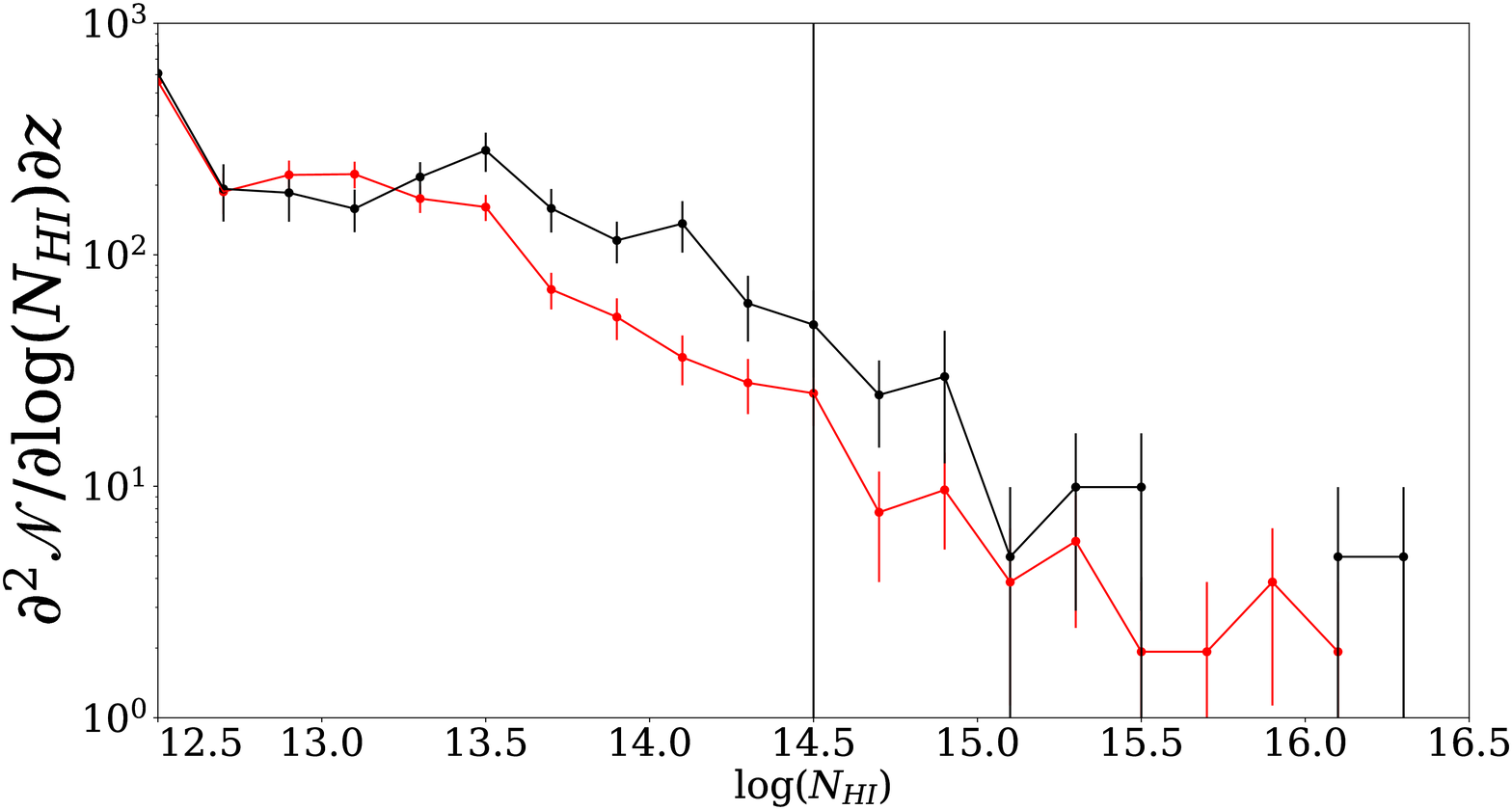}
    \caption{Column density distribution (CDD) of \textit{HST}/COS Ly~$\alpha$ absorbers in walls (black) and voids (red). This function, the number of absorbers per logarithmic interval of column density per unit redshift pathlength, is shallower in walls because they contain relatively more high column density systems. For the sake of comparison, the upper limit of the range of column density examined in Figure 2 of \citet{2017ApJ...845...47T} is marked with a vertical bar.}
    \label{fig:cdd}
\end{figure}

A more detailed probe of environmental differences in the IGM is the column density distribution (CDD) of Ly~$\alpha$ absorbers, the number of absorbers per interval of column density per unit redshift pathlength, in voids and walls. Figure \ref{fig:cdd} is a log-log plot of the CDD of Ly~$\alpha$ absorbers in voids and walls (see Figure 5 in \citetalias{2016ApJ...817..111D} for the CDD of the entire set of Ly~$\alpha$ absorbers). The CDD is computed as in \citet{2008ApJ...679..194D}. We correct the effective redshift path length $\Delta z$ for void and wall absorbers using sight line randoms to estimate the fraction of sight lines in voids and walls, respectively 72 and 28 per cent; as discussed above, the majority of absorbers are in voids.
Thus normalized, the CDDs quantify the frequency of absorbers per unit pathlength in the separate environments.
Note that at small column densities observations of both wall and void absorbers are limited by the the sensitivity limits of \textit{HST}/COS. 
Correction for this limit is important because simulations suggest that most discrete Ly~$\alpha$ absorbers are low-N\textsubscript{HI} \citep{2010MNRAS.408.2051D}. 
Completeness correction factors at the lower range of column density (identical for void and wall absorbers) are based on a fifth order polynomial fit to redshift pathlengths for bins of column density in Table 5 of \citetalias{2016ApJ...817..111D}. 

The steepness of the CDD is quantified by $\beta$, where $\partial \mathscr{N} / \partial N \partial z \propto N^{-\beta} $ (that is, the slope in Figure \ref{fig:cdd} plus one) \citep{Dave_2001}. Over the full range of column density observed, we find $\beta =$ 1.72 $\pm$ 0.03 for void absorbers, steeper than in walls, where $\beta=$1.54 $\pm$ 0.05. Compared to voids, walls contain a higher ratio of high to low column density systems. The shallower slope of the wall CDD is likely connected to the fact that higher column density systems are more closely associated with galaxies, which by definition are primarily found in walls.  For the combination of void and wall absorbers, $\beta =$ 1.65 $\pm$ 0.02. This is matches the value obtained by \citetalias{2016ApJ...817..111D} for the first edition of their IGM absorber catalog. \citet{2010MNRAS.408.2051D} simulated the CDD of Ly~$\alpha$ absorbers for HST/COS and predicted a somewhat higher value, $\beta =$ 1.70 $\pm$ 0.1.The voids dominate the statistics of the combined distribution because they contain the majority (about 70 per cent) of absorbers.

Our observations of the void and wall CDD contrast with simulation results of \citet{2017ApJ...845...47T}. They compared the CDDs of Ly-$\alpha$ absorbers found in hydrodynamic simulations of two environments: an underdense region centered on a void and an overdense region centered on a galaxy cluster. They found a steeper CDD in the overdense region than the underdense region over 12.5~$<$~$\log$N\textsubscript{HI}~$<$14.5 (to the left of the vertical line in our Figure \ref{fig:cdd}), primarily because the overdense region had fewer high column density absorbers. 
If we restrict our computation of $\beta$ to the more limited range of column density that they examined, we find values for $\beta$ in voids and walls (1.70 $\pm$ 0.06 and 1.63 $\pm$ 0.11 respectively) that are similar to our findings for the full range of column density; the observed wall CDD is shallower than the void CDD. 
The different findings in these simulations and our observations may be affected by differences in classification of large-scale environment. By including small voids in our definition of walls (recall that VoidFinder imposes a minimum void radius of $R_v=10h^{-1}$~Mpc), our wall sample may include some absorbers in underdense regions. 
Another possibility is that this disagreement indicates that the two regions examined in the simulations are not representative of typical overdense and underdense volumes in the low-redshift universe. Perhaps the environment of the $+1.8\sigma$ "cluster" region in the simulation suppresses absorber abundance more than the  typical overdense environment sampled by our "wall" regions. \citet{2017ApJ...845...47T} find that the underlying IGM clouds in their cluster simulation region are typically at least a factor of 1.5 hotter than in the void region. In the temperature regime $T\sim 10^{4.5}$ K, such a temperature difference would manifest as a roughly 4 km~s$^{-1}$ increase in the $b$ parameter (for purely thermal broadening, $b=12.8(T/10^4\, \mathrm{K})^{1/2}$ \ km~s$^{-1}$), but our measurement of Doppler $b$ parameters for wall and void absorbers does not yield evidence of such a shift (note again that the measured $b$ parameters are also substantially broadened by the spectroscopic resolution of \textit{HST}/COS).
Our CDD is less affected by cosmic variance in the void and wall environments, because we examined Ly~$\alpha$ absorbers in a much larger volume, including hundreds of voids.

Next we test for variation with column density of the location of absorbers within voids.
We might expect lower column density absorbers to be more prevalent in the deep interiors of voids, where the IGM is isolated from shock heating from structure formation in dense regions. To test for that effect, we split the 392 void absorbers into quartiles of column density and perform a very simple test: What fraction of absorbers are found at less than half the void radius?
For each quartile, we count the number of Ly~$\alpha$ absorbers in the inner eighth of void volume, N\textsubscript{obs} (r/ R\textsubscript v < 1/2). The total number of objects N in voids is compared with N\textsubscript{obs} (r/ R\textsubscript v < 1/2), the number inside the inner eighth of void volume, and N\textsubscript{ran} (r/ R\textsubscript v < 1/2), the number expected from Poisson statistics.

Thus, we find suggestive evidence that low-N\textsubscript{HI} absorbers are more centrally concentrated in voids and that the highest column density absorbers are less centrally concentrated, as summarized in Table \ref{tab:stattest}. In other words, there is a trend for lower column density absorbers to be overrepresented in the central regions of voids. 
This accords with the observation that low-N\textsubscript{HI} absorbers are the least strongly correlated with galaxies, which are most sparse in the void interiors.

\begin{table*}
	\centering
	\caption{A summary of our simple statistical test of the central concentration of absorbers and galaxies in voids. The total number of objects N in voids is compared with N\textsubscript{obs} (r/ R\textsubscript v < 1/2), the number inside the inner eighth of void volume (within half the void radius), and N\textsubscript{ran} (r/ R\textsubscript v < 1/2), the number expected from the fractions of sight line randoms and survey randoms inside a scaled voidcentric distance of 1/2. Our set of 36 AGN sight lines samples a higher fraction of the interior void volume. This is why the expected fraction of void absorbers in each quartile is higher than 1/8. Void galaxies are significantly less centrally concentrated than expected for a random distribution. For void absorbers we observe that the lowest column density quartile Q1 is more centrally concentrated, with the lowest quartile of column density being overrepresented in void centers compared to a random distribution. Conversely, the highest column density absorbers are underrepresented in void centers.}
	\label{tab:stattest}
	\begin{tabular}{ | c | c | c | c | c | c | c |}
		 & Gal. & Ly~$\alpha$ & Q1 & Q2 & Q3 & Q4\\
		\hline
    		N & 26,951 & 392 & 98 & 98 & 98 & 98\\ 
    		N\textsubscript{obs} (r/ R\textsubscript v < 1/2) & 1117 & 72 & 31 & 19 & 14 & 8\\ 
    		N\textsubscript{ran} (r/ R\textsubscript v < 1/2) & 3369 $\pm$ 58 & 84 $\pm$ 20 & 21 $\pm$ 5 & 21 $\pm$ 5 & 21 $\pm$ 5 & 21 $\pm$ 5\\
		\hline
	\end{tabular}
\end{table*}

\subsection{Galaxy Association of Void Ly \texorpdfstring{$\alpha$}{} Absorbers}
To probe interactions between galaxies, the CGM, and the IGM, we find the distance, d\textsubscript{g}, of each Ly~$\alpha$ absorber to the nearest SDSS DR7 galaxy in both volume-limited and apparent-magnitude limited samples. Following the procedure in \citet{2018ApJS..237...11K}, we compute the velocity separation $\Delta v$ between a galaxy and absorber to find distance along the line of sight 
\[ D_z = \begin{cases} 
      \frac{| \Delta v | \ - \ 400 \ km \ s^{-1}}{H(z)} & \Delta v \ > \ 400 \ km \ s^{-1} \\
      0 & \Delta v \leq \ 400 \ km \ s^{-1} 
   \end{cases}
\]
where $H(z) = 100 h~km~s^{-1}~Mpc^{-1}$. For low velocity separation, small line-of-sight distances are collapsed to zero. This is a correction for small-scale redshift space distortions along the line-of-sight caused by gravitational motions within systems of galaxies. The impact parameter of the galaxy with respect to the AGN sight line and $D_z$ are added in quadrature to obtain the three dimensional distance $d\textsubscript{g}$.

\begin{figure}
	\includegraphics[width=\columnwidth]{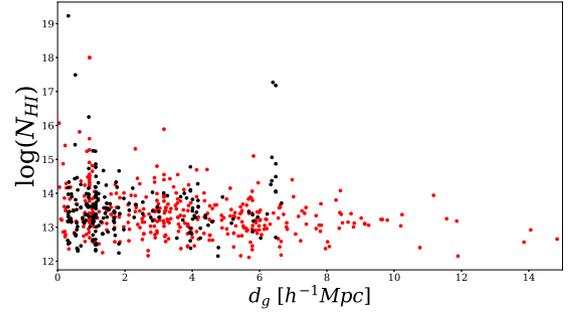}
    \caption{Scatterplot of atomic hydrogen column density $\log{N_{HI}}$ vs. distance to nearest galaxy d\textsubscript{g} in the volume-limited sample (0.003~$\leq$~z~$\leq$~0.107, M\textsuperscript{r}~$<$~-20.09) of SDSS DR7 for the 605 \textit{HST}/COS absorbers in SDSS DR7. Wall absorbers are black, void absorbers red.}
    \label{fig:NHIdgal}
\end{figure}

\begin{figure}
	\includegraphics[width=\columnwidth]{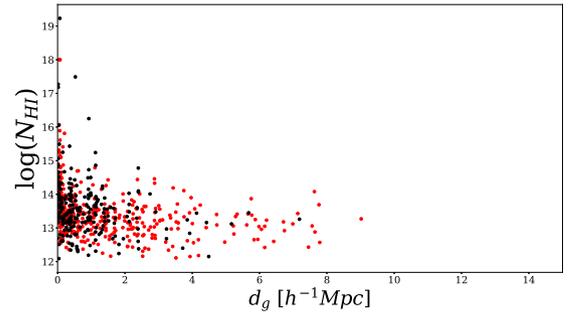}
    \caption{Scatterplot of atomic hydrogen column density $\log{N_{HI}}$ vs. distance to nearest galaxy, d\textsubscript{g}, in the apparent-magnitude limited ($<17.77$) galaxy sample in SDSS DR7 ($0.003 \leq z \leq 0.107$) for the 605 \textit{HST}/COS absorbers in SDSS DR7. Wall absorbers are black, void absorbers red.}
    \label{fig:NHIdgal_all}
\end{figure}

Figures \ref{fig:NHIdgal} and \ref{fig:NHIdgal_all} show distance to nearest galaxy, d\textsubscript{g}, versus column density, N\textsubscript{HI}, for void (red) and wall absorbers (black), comparing respectively with volume-limited and apparent-magnitude limited samples of SDSS DR7. As expected, the denser apparent-magnitude limited sample yields closer matches. Previous studies (e.g. \citealt{2002ApJ...565..720P,2006ApJ...641..217S,2018ApJS..237...11K}) have generally found that Ly~$\alpha$ absorbers at higher d\textsubscript{g} have lower equivalent width or column density. Our results are similar, in that higher-N\textsubscript{HI} absorbers tend to be more closely associated with galaxies. 
The large-scale environment of Ly$\alpha$ absorbers in walls or voids does not noticeably affect the distribution of d\textsubscript{g}, apart from the fact that void absorbers are considerably more prevalent at higher d\textsubscript{g}, because they occupy regions of lower galaxy density. 
In Figure \ref{fig:NHIdgal_all} we see a population of void absorbers that are close to galaxies (d\textsubscript{g}~$<$~1~h\textsuperscript{-1}~Mpc); these matches tend to occur close to the void edges. A relatively small number of wall absorbers are also relatively isolated from bright galaxies; these occupy voids that are below the cutoff radius of the VoidFinder algorithm, 10~h\textsuperscript{-1}~Mpc (in other words, a less stringent definition of voids would have caused those to be labeled "void absorbers").

\section{Discussion}
\label{sec:discussion}
The spatial distribution of \textit{HST}/COS Ly$\alpha$ absorbers inside voids shares characteristics with both a uniform random distribution and the strongly clustered galaxy distribution. The void density profile of Ly$\alpha$ absorbers as a function of scaled voidcentric distance, shown in Figure \ref{fig:rho_rnorm}, appears consistent with a uniform distribution; the prominent density spike in the lowest bin is not statistically significant, as indicated by the size of the errorbar. On the other hand, there is a significant density enhancement of Ly$\alpha$ absorbers close to the void-wall boundary, as shown in Figure \ref{fig:rho_dwall}, which plots the void density profile of Ly$\alpha$ absorbers as a function of distance to cosmic walls. This density enhancement near void boundaries shows that Ly$\alpha$ absorbers are not randomly distributed inside voids. A similar, possibly related rise in density appears near void boundaries in the density profile of galaxies shown in Figure \ref{fig:rho_dwall}. 

We conclude that there are two populations of Ly$\alpha$ absorbers inside voids: one with a spatial distribution that appears nearly random, and another associated with galaxies near the edges of voids. This interpretation is supported by our analysis of the cumulative distribution function of Ly$\alpha$ absorbers inside voids. Their cumulative distribution is close to random, but slightly skewed toward the void edges. As seen in Figure \ref{fig:NHIdgal_all}, only lower column density absorbers ($\log{N_{HI}}<15$) are far from any galaxies. This isolated population of Ly$\alpha$ absorbers is mainly found in voids. There are also some wall absorbers that are far from any galaxies, but such absorbers reside in underdense regions that are too small to be classified as voids by the VoidFinder algorithm (the maximal sphere radius cannot be less than 10~h\textsuperscript{-1}~Mpc). By contrast, the absorbers associated with galaxies include both low and high column density systems.

Previous research has also suggested there may be two populations of IGM clouds detected in Ly$\alpha$ absorption. \citet{2021ApJ...912....9W} found that high column density absorbers (N\textsubscript{HI} $>$ 10\textsuperscript{14} cm\textsuperscript{-2}) are usually close enough to be physically associated with a galaxy (within two virial radii of a galaxy). Weaker Ly~$\alpha$ absorbers were more widely distributed, probably because they tend to be associated with extended large-scale structure but not particular galaxies or galaxy groups. \citet{2012MNRAS.425..245T}, a study of \textit{HST}/STIS absorbers in SDSS DR7, proposed that there are three populations of Ly$\alpha$ absorbers associated with galaxies, overdense large-scale structure, and underdense large-scale structure. They noted that absorbers in underdense large-scale structure, or the interior volume of voids, may not be randomly distributed. 

Ly~$\alpha$ absorption primarily traces a diffuse phase of the IGM, defined by \citep{1996ApJ...469..480K} as being below a threshold temperature (T$<$10\textsuperscript{5}) and threshold density,
\begin{equation}
    \delta_{th} = 6 \pi^2 (1+0.4093(\frac{1}{f_{\Omega}} -1)^{0.9052})
\end{equation}
where 
\begin{equation}
    f_{\Omega} = \frac{\Omega_m (1+z)^3}{\Omega_m (1+z)^3 + (1 - \Omega_m - \Omega_{\Lambda})(1+z)^2 + \Omega_{\Lambda}}
\end{equation}
Cosmological simulations \citep{2010MNRAS.408.2051D, 2021arXiv210913378R} suggest that this diffuse phase constitutes about 40 per cent of the IGM. Hotter, more ionized phases are weak in Ly~$\alpha$, but may be traced by the O VI absorption line. Thus the spatial distribution of Ly~$\alpha$ absorbers depends not only on the baryon distribution, but also the ionisation fraction of intergalactic hydrogen. Environmental conditions in the interiors of voids, namely isolation from the effects of shock-heating from structure formation and adiabatic super-Hubble expansion, cool and lower the ionisation fraction of intergalactic hydrogen.  Hydrodynamic simulations by \citet{2011ApJ...741...99C} show that shock heating of halo gas in dense wall regions prevents most of it from cooling within the current age of the universe. This likely explains why voids have more Ly~$\alpha$ absorbers than walls, despite the fact that they contain less gas by mass.

We find a significant over-representation of \textit{HST}/COS absorbers in the lowest quartile of column density inside the interior eighth of void volume (compared to a random distribution), which suggests that there could be a special population of low column density absorbers created by cold and diffuse hydrogen near the centers of voids. This population could contribute to the peak in the density profile of Ly$\alpha$ absorbers near the void center, in the lowest bin of Figure \ref{fig:rho_rnorm}, although as noted previously, the peak is not statistically significant. Results from cosmological hydrodynamic simulations also suggest that cold, neutral hydrogen gas is preferentially found near void centres, where there is no shock-heating from structure formation. \citet{2011ApJ...735..132K}, a study of galaxies in the \textit{Enzo} simulation \citep{1999CSE.....1b..46B}, found that low-luminosity dwarf galaxies are over-represented near the centre of the void region. A more recent cosmological simulation using over a thousand voids by \citet{2021arXiv210913378R} found that most intergalactic hydrogen in the interiors of voids exists in a diffuse, low temperature phase (T$<$10\textsuperscript{15} K), which is strong in Ly~$\alpha$ absorption.

We have focused our examination on Ly~$\alpha$ absorption systems, but note that metal absorption components present in \textit{HST}/COS AGN spectra can provide a fuller picture of the gas conditions in voids. Since intergalactic hydrogen is highly ionised, it is mostly undetectable in FUV absorption. \citet{2019ApJS..240...15S}, a spectroscopic study of gas associated with galaxy groups in the FUV, failed to detect a diffuse intergroup medium that would complete the baryon census. \citet{2018Natur.558..406N} claimed detection of these missing baryons through x-ray spectroscopy. The missing baryons are a hot phase of the WHIM ($T=10^{5.7}-10^{6.3} \ K$) located in large galaxy overdensities. Complex feedback between the IGM, CGM, and galaxies shape their evolution. Matter circulated from the IGM into galaxies fuels star formation \citep{2008ApJ...674..151E, 2009ApJ...696.1543P, 2010MNRAS.407.2091G}. Supernovae and AGN feedback cause gas outflow from galaxies into the CGM and ultimately IGM, altering their primordial composition by introducing metals (e.g. \citealt{2001ApJ...555...92M, 2014MNRAS.444L.105P,2006MNRAS.370..645B}). We note that three O VI components, which trace WHIM, fall within the footprint and redshift limit of our SDSS DR7 void catalog. Two of these are inside voids, and one is well within the void interior, at a scaled voidcentric distance of less than a half. 

\section{Conclusions}
\label{sec:conclusion}
We characterize the spatial distribution and intrinsic properties of \textit{HST}/COS Ly~$\alpha$ absorbers within voids detected in the nearby volume mapped by SDSS DR7. These voids fill 68 per cent of the survey volume, and contain a comparable fraction of absorbers, but only a fifth of the galaxies. The void density profile of all Ly~$\alpha$ absorbers as a function of scaled voidcentric distance appears consistent with a uniform distribution in the interiors of voids. Similarly, the cumulative distribution function of Ly~$\alpha$ absorbers inside voids shows they are similar to a random distribution, but have a slight tendency to reside close to the edges of voids. In detail, the spatial distribution of absorbers inside voids depends on HI column density. Absorbers in the lowest quartile of HI column density are more centrally concentrated in the deep interiors of voids, whereas absorbers in the highest quartile of column density are skewed toward the void wall boundaries. Near the void-wall boundary, we find a slight but statistically significant enhancement in Ly~$\alpha$ absorber density, in the same region where galaxy density prominently rises to form the walls that bound voids. We interpret these findings in terms of two populations of Ly~$\alpha$ absorbers in the low-redshift universe. One is associated with galaxies and covers the full range of column density. This population is likely to include gas that is in the diffuse IGM phase as well as some WHIM as defined by \citet{2010MNRAS.408.2051D}. The other population of absorbers is more uniformly distributed, consists only of lower column density systems ($\log$N\textsubscript{HI}~$<$~14.5), and is likely to be purely in the diffuse IGM phase.

\section*{Data Availability}
The data underlying this article are available at \url{https://archive.stsci.edu/prepds/igm/}.

\bibliographystyle{mnras}
\bibliography{refs} 

\begin{thebibliography}{}
\makeatletter
\relax
\def\mn@urlcharsother{\let\do\@makeother \do\$\do\&\do\#\do\^\do\_\do\%\do\~}
\def\mn@doi{\begingroup\mn@urlcharsother \@ifnextchar [ {\mn@doi@}
  {\mn@doi@[]}}
\def\mn@doi@[#1]#2{\def\@tempa{#1}\ifx\@tempa\@empty \href
  {http://dx.doi.org/#2} {doi:#2}\else \href {http://dx.doi.org/#2} {#1}\fi
  \endgroup}
\def\mn@eprint#1#2{\mn@eprint@#1:#2::\@nil}
\def\mn@eprint@arXiv#1{\href {http://arxiv.org/abs/#1} {{\tt arXiv:#1}}}
\def\mn@eprint@dblp#1{\href {http://dblp.uni-trier.de/rec/bibtex/#1.xml}
  {dblp:#1}}
\def\mn@eprint@#1:#2:#3:#4\@nil{\def\@tempa {#1}\def\@tempb {#2}\def\@tempc
  {#3}\ifx \@tempc \@empty \let \@tempc \@tempb \let \@tempb \@tempa \fi \ifx
  \@tempb \@empty \def\@tempb {arXiv}\fi \@ifundefined
  {mn@eprint@\@tempb}{\@tempb:\@tempc}{\expandafter \expandafter \csname
  mn@eprint@\@tempb\endcsname \expandafter{\@tempc}}}

\bibitem[\protect\citeauthoryear{{Abazajian} et~al.,}{{Abazajian}
  et~al.}{2009}]{2009ApJS..182..543A}
{Abazajian} K.~N.,  et~al., 2009, \mn@doi [\apjs]
  {10.1088/0067-0049/182/2/543}, \href
  {http://adsabs.harvard.edu/abs/2009ApJS..182..543A} {182, 543}

\bibitem[\protect\citeauthoryear{{Bahcall}, {Jannuzi}, {Schneider}, {Hartig},
  {Bohlin}  \& {Junkkarinen}}{{Bahcall} et~al.}{1991}]{1991ApJ...377L...5B}
{Bahcall} J.~N.,  {Jannuzi} B.~T.,  {Schneider} D.~P.,  {Hartig} G.~F.,
  {Bohlin} R.,   {Junkkarinen} V.,  1991, \mn@doi [\apjl] {10.1086/186103},
  \href {http://adsabs.harvard.edu/abs/1991ApJ...377L...5B} {377, L5}

\bibitem[\protect\citeauthoryear{{Bahcall} et~al.,}{{Bahcall}
  et~al.}{1996}]{1996ApJ...457...19B}
{Bahcall} J.~N.,  et~al., 1996, \mn@doi [\apj] {10.1086/176709}, \href
  {http://adsabs.harvard.edu/abs/1996ApJ...457...19B} {457, 19}

\bibitem[\protect\citeauthoryear{{Blanton}, {Kazin}, {Muna}, {Weaver}  \&
  {Price-Whelan}}{{Blanton} et~al.}{2011}]{2011AJ....142...31B}
{Blanton} M.~R.,  {Kazin} E.,  {Muna} D.,  {Weaver} B.~A.,   {Price-Whelan} A.,
   2011, \mn@doi [\aj] {10.1088/0004-6256/142/1/31}, \href
  {https://ui.adsabs.harvard.edu/abs/2011AJ....142...31B} {142, 31}

\bibitem[\protect\citeauthoryear{{Boggess} et~al.,}{{Boggess}
  et~al.}{1978}]{1978Natur.275..372B}
{Boggess} A.,  et~al., 1978, \mn@doi [\nat] {10.1038/275372a0}, \href
  {https://ui.adsabs.harvard.edu/abs/1978Natur.275..372B} {275, 372}

\bibitem[\protect\citeauthoryear{{Bower}, {Benson}, {Malbon}, {Helly}, {Frenk},
  {Baugh}, {Cole}  \& {Lacey}}{{Bower} et~al.}{2006}]{2006MNRAS.370..645B}
{Bower} R.~G.,  {Benson} A.~J.,  {Malbon} R.,  {Helly} J.~C.,  {Frenk} C.~S.,
  {Baugh} C.~M.,  {Cole} S.,   {Lacey} C.~G.,  2006, \mn@doi [\mnras]
  {10.1111/j.1365-2966.2006.10519.x}, \href
  {https://ui.adsabs.harvard.edu/abs/2006MNRAS.370..645B} {370, 645}

\bibitem[\protect\citeauthoryear{{Bryan}}{{Bryan}}{1999}]{1999CSE.....1b..46B}
{Bryan} G.~L.,  1999, \mn@doi [Computing in Science and Engineering]
  {10.1109/5992.753046}, \href
  {https://ui.adsabs.harvard.edu/abs/1999CSE.....1b..46B} {1, 46}

\bibitem[\protect\citeauthoryear{{Carswell} \& {Rees}}{{Carswell} \&
  {Rees}}{1987}]{1987MNRAS.224P..13C}
{Carswell} R.~F.,  {Rees} M.~J.,  1987, \mn@doi [\mnras]
  {10.1093/mnras/224.1.13P}, \href
  {http://adsabs.harvard.edu/abs/1987MNRAS.224P..13C} {224, 13P}

\bibitem[\protect\citeauthoryear{{Cen}}{{Cen}}{2011}]{2011ApJ...741...99C}
{Cen} R.,  2011, \mn@doi [\apj] {10.1088/0004-637X/741/2/99}, \href
  {https://ui.adsabs.harvard.edu/\#abs/2011ApJ...741...99C} {741, 99}

\bibitem[\protect\citeauthoryear{{Cen} \& {Ostriker}}{{Cen} \&
  {Ostriker}}{1999}]{1999ApJ...514....1C}
{Cen} R.,  {Ostriker} J.~P.,  1999, \mn@doi [\apj] {10.1086/306949}, \href
  {https://ui.adsabs.harvard.edu/\#abs/1999ApJ...514....1C} {514, 1}

\bibitem[\protect\citeauthoryear{{Cen}, {Miralda-Escud{\'e}}, {Ostriker}  \&
  {Rauch}}{{Cen} et~al.}{1994}]{1994ApJ...437L...9C}
{Cen} R.,  {Miralda-Escud{\'e}} J.,  {Ostriker} J.~P.,   {Rauch} M.,  1994,
  \mn@doi [\apjl] {10.1086/187670}, \href
  {http://adsabs.harvard.edu/abs/1994ApJ...437L...9C} {437, L9}

\bibitem[\protect\citeauthoryear{{Colberg}}{{Colberg}}{2004}]{2004ogci.conf...51C}
{Colberg} J.~M.,  2004, in {Diaferio} A.,  ed., IAU Colloq. 195: Outskirts of
  Galaxy Clusters: Intense Life in the Suburbs. pp 51--57,
  \mn@doi{10.1017/S1743921304000122}

\bibitem[\protect\citeauthoryear{{Croft}, {Weinberg}, {Katz}  \&
  {Hernquist}}{{Croft} et~al.}{1998}]{1998ApJ...495...44C}
{Croft} R. A.~C.,  {Weinberg} D.~H.,  {Katz} N.,   {Hernquist} L.,  1998,
  \mn@doi [\apj] {10.1086/305289}, \href
  {https://ui.adsabs.harvard.edu/\#abs/1998ApJ...495...44C} {495, 44}

\bibitem[\protect\citeauthoryear{{Croft}, {Weinberg}, {Pettini}, {Hernquist}
  \& {Katz}}{{Croft} et~al.}{1999}]{1999ApJ...520....1C}
{Croft} R. A.~C.,  {Weinberg} D.~H.,  {Pettini} M.,  {Hernquist} L.,   {Katz}
  N.,  1999, \mn@doi [\apj] {10.1086/307438}, \href
  {https://ui.adsabs.harvard.edu/\#abs/1999ApJ...520....1C} {520, 1}

\bibitem[\protect\citeauthoryear{{Croft}, {Weinberg}, {Bolte}, {Burles},
  {Hernquist}, {Katz}, {Kirkman}  \& {Tytler}}{{Croft}
  et~al.}{2002}]{2002ApJ...581...20C}
{Croft} R. A.~C.,  {Weinberg} D.~H.,  {Bolte} M.,  {Burles} S.,  {Hernquist}
  L.,  {Katz} N.,  {Kirkman} D.,   {Tytler} D.,  2002, \mn@doi [\apj]
  {10.1086/344099}, \href
  {https://ui.adsabs.harvard.edu/\#abs/2002ApJ...581...20C} {581, 20}

\bibitem[\protect\citeauthoryear{{Danforth} \& {Shull}}{{Danforth} \&
  {Shull}}{2008}]{2008ApJ...679..194D}
{Danforth} C.~W.,  {Shull} J.~M.,  2008, \mn@doi [\apj] {10.1086/587127}, \href
  {http://adsabs.harvard.edu/abs/2008ApJ...679..194D} {679, 194}

\bibitem[\protect\citeauthoryear{{Danforth} et~al.,}{{Danforth}
  et~al.}{2016}]{2016ApJ...817..111D}
{Danforth} C.~W.,  et~al., 2016, \mn@doi [\apj] {10.3847/0004-637X/817/2/111},
  \href {http://adsabs.harvard.edu/abs/2016ApJ...817..111D} {817, 111}

\bibitem[\protect\citeauthoryear{Dave \& Tripp}{Dave \&
  Tripp}{2001}]{Dave_2001}
Dave R.,  Tripp T.~M.,  2001, \mn@doi [The Astrophysical Journal]
  {10.1086/320977}, 553, 528

\bibitem[\protect\citeauthoryear{{Dav{\'e}}, {Hernquist}, {Katz}  \&
  {Weinberg}}{{Dav{\'e}} et~al.}{1999}]{1999ApJ...511..521D}
{Dav{\'e}} R.,  {Hernquist} L.,  {Katz} N.,   {Weinberg} D.~H.,  1999, \mn@doi
  [\apj] {10.1086/306722}, \href
  {https://ui.adsabs.harvard.edu/\#abs/1999ApJ...511..521D} {511, 521}

\bibitem[\protect\citeauthoryear{{Dav{\'e}}, {Oppenheimer}, {Katz}, {Kollmeier}
   \& {Weinberg}}{{Dav{\'e}} et~al.}{2010}]{2010MNRAS.408.2051D}
{Dav{\'e}} R.,  {Oppenheimer} B.~D.,  {Katz} N.,  {Kollmeier} J.~A.,
  {Weinberg} D.~H.,  2010, \mn@doi [\mnras] {10.1111/j.1365-2966.2010.17279.x},
  \href {http://adsabs.harvard.edu/abs/2010MNRAS.408.2051D} {408, 2051}

\bibitem[\protect\citeauthoryear{{Douglass}, {Veyrat}  \& {BenZvi}}{{Douglass}
  et~al.}{2022}]{2022arXiv220201226D}
{Douglass} K.~A.,  {Veyrat} D.,   {BenZvi} S.,  2022, arXiv e-prints, \href
  {https://ui.adsabs.harvard.edu/abs/2022arXiv220201226D} {p. arXiv:2202.01226}

\bibitem[\protect\citeauthoryear{{El-Ad} \& {Piran}}{{El-Ad} \&
  {Piran}}{1997}]{1997ApJ...491..421E}
{El-Ad} H.,  {Piran} T.,  1997, \mn@doi [\apj] {10.1086/304973}, \href
  {https://ui.adsabs.harvard.edu/abs/1997ApJ...491..421E} {491, 421}

\bibitem[\protect\citeauthoryear{{Erb}}{{Erb}}{2008}]{2008ApJ...674..151E}
{Erb} D.~K.,  2008, \mn@doi [\apj] {10.1086/524727}, \href
  {https://ui.adsabs.harvard.edu/abs/2008ApJ...674..151E} {674, 151}

\bibitem[\protect\citeauthoryear{{Frank} et~al.,}{{Frank}
  et~al.}{2012}]{2012MNRAS.420.1731F}
{Frank} S.,  et~al., 2012, \mn@doi [\mnras] {10.1111/j.1365-2966.2011.20172.x},
  \href {http://adsabs.harvard.edu/abs/2012MNRAS.420.1731F} {420, 1731}

\bibitem[\protect\citeauthoryear{{Fukugita}, {Ichikawa}, {Gunn}, {Doi},
  {Shimasaku}  \& {Schneider}}{{Fukugita} et~al.}{1996}]{1996AJ....111.1748F}
{Fukugita} M.,  {Ichikawa} T.,  {Gunn} J.~E.,  {Doi} M.,  {Shimasaku} K.,
  {Schneider} D.~P.,  1996, \mn@doi [\aj] {10.1086/117915}, \href
  {http://adsabs.harvard.edu/abs/1996AJ....111.1748F} {111, 1748}

\bibitem[\protect\citeauthoryear{{Genzel} et~al.,}{{Genzel}
  et~al.}{2010}]{2010MNRAS.407.2091G}
{Genzel} R.,  et~al., 2010, \mn@doi [\mnras]
  {10.1111/j.1365-2966.2010.16969.x}, \href
  {https://ui.adsabs.harvard.edu/abs/2010MNRAS.407.2091G} {407, 2091}

\bibitem[\protect\citeauthoryear{{Grogin} \& {Geller}}{{Grogin} \&
  {Geller}}{1998}]{1998ApJ...505..506G}
{Grogin} N.~A.,  {Geller} M.~J.,  1998, \mn@doi [\apj] {10.1086/306208}, \href
  {http://adsabs.harvard.edu/abs/1998ApJ...505..506G} {505, 506}

\bibitem[\protect\citeauthoryear{{Gunn} \& {Peterson}}{{Gunn} \&
  {Peterson}}{1965}]{1965ApJ...142.1633G}
{Gunn} J.~E.,  {Peterson} B.~A.,  1965, \mn@doi [\apj] {10.1086/148444}, \href
  {http://adsabs.harvard.edu/abs/1965ApJ...142.1633G} {142, 1633}

\bibitem[\protect\citeauthoryear{{Gunn} et~al.,}{{Gunn}
  et~al.}{1998}]{1998AJ....116.3040G}
{Gunn} J.~E.,  et~al., 1998, \mn@doi [\aj] {10.1086/300645}, \href
  {http://adsabs.harvard.edu/abs/1998AJ....116.3040G} {116, 3040}

\bibitem[\protect\citeauthoryear{{Haardt} \& {Madau}}{{Haardt} \&
  {Madau}}{2001}]{2001cghr.confE..64H}
{Haardt} F.,  {Madau} P.,  2001, in {Neumann} D.~M.,  {Tran} J.~T.~V.,  eds,
  Clusters of Galaxies and the High Redshift Universe Observed in X-rays. p.~64
  (\mn@eprint {arXiv} {astro-ph/0106018})

\bibitem[\protect\citeauthoryear{{Hernquist}, {Katz}, {Weinberg}  \&
  {Miralda-Escud{\'e}}}{{Hernquist} et~al.}{1996}]{1996ApJ...457L..51H}
{Hernquist} L.,  {Katz} N.,  {Weinberg} D.~H.,   {Miralda-Escud{\'e}} J.,
  1996, \mn@doi [\apjl] {10.1086/309899}, \href
  {http://adsabs.harvard.edu/abs/1996ApJ...457L..51H} {457, L51}

\bibitem[\protect\citeauthoryear{{Hoyle} \& {Vogeley}}{{Hoyle} \&
  {Vogeley}}{2002}]{2002ApJ...566..641H}
{Hoyle} F.,  {Vogeley} M.~S.,  2002, \mn@doi [\apj] {10.1086/338340}, \href
  {https://ui.adsabs.harvard.edu/\#abs/2002ApJ...566..641H} {566, 641}

\bibitem[\protect\citeauthoryear{{Hoyle} \& {Vogeley}}{{Hoyle} \&
  {Vogeley}}{2004}]{2004ApJ...607..751H}
{Hoyle} F.,  {Vogeley} M.~S.,  2004, \mn@doi [\apj] {10.1086/386279}, \href
  {https://ui.adsabs.harvard.edu/\#abs/2004ApJ...607..751H} {607, 751}

\bibitem[\protect\citeauthoryear{{Hu}, {Kim}, {Cowie}, {Songaila}  \&
  {Rauch}}{{Hu} et~al.}{1995}]{1995AJ....110.1526H}
{Hu} E.~M.,  {Kim} T.-S.,  {Cowie} L.~L.,  {Songaila} A.,   {Rauch} M.,  1995,
  \mn@doi [\aj] {10.1086/117625}, \href
  {https://ui.adsabs.harvard.edu/\#abs/1995AJ....110.1526H} {110, 1526}

\bibitem[\protect\citeauthoryear{{Jarosik} et~al.,}{{Jarosik}
  et~al.}{2011}]{2011ApJS..192...14J}
{Jarosik} N.,  et~al., 2011, \mn@doi [The Astrophysical Journal Supplement
  Series] {10.1088/0067-0049/192/2/14}, \href
  {https://ui.adsabs.harvard.edu/\#abs/2011ApJS..192...14J} {192, 14}

\bibitem[\protect\citeauthoryear{{Keeney} et~al.,}{{Keeney}
  et~al.}{2018}]{2018ApJS..237...11K}
{Keeney} B.~A.,  et~al., 2018, \mn@doi [The Astrophysical Journal Supplement
  Series] {10.3847/1538-4365/aac727}, \href
  {https://ui.adsabs.harvard.edu/\#abs/2018ApJS..237...11K} {237, 11}

\bibitem[\protect\citeauthoryear{{Kitayama} \& {Suto}}{{Kitayama} \&
  {Suto}}{1996}]{1996ApJ...469..480K}
{Kitayama} T.,  {Suto} Y.,  1996, \mn@doi [\apj] {10.1086/177797}, \href
  {https://ui.adsabs.harvard.edu/abs/1996ApJ...469..480K} {469, 480}

\bibitem[\protect\citeauthoryear{{Kreckel}, {Joung}  \& {Cen}}{{Kreckel}
  et~al.}{2011}]{2011ApJ...735..132K}
{Kreckel} K.,  {Joung} M.~R.,   {Cen} R.,  2011, \mn@doi [\apj]
  {10.1088/0004-637X/735/2/132}, \href
  {https://ui.adsabs.harvard.edu/abs/2011ApJ...735..132K} {735, 132}

\bibitem[\protect\citeauthoryear{{Labatie}, {Starck}, {Lachi{\`e}ze-Rey}  \&
  {Arnalte-Mur}}{{Labatie} et~al.}{2010}]{2010arXiv1009.1232L}
{Labatie} A.,  {Starck} J.-L.,  {Lachi{\`e}ze-Rey} M.,   {Arnalte-Mur} P.,
  2010, arXiv e-prints, \href
  {https://ui.adsabs.harvard.edu/\#abs/2010arXiv1009.1232L} {p.
  arXiv:1009.1232}

\bibitem[\protect\citeauthoryear{{Li}, {Kauffmann}, {Jing}, {White},
  {B{\"o}rner}  \& {Cheng}}{{Li} et~al.}{2006}]{2006MNRAS.368...21L}
{Li} C.,  {Kauffmann} G.,  {Jing} Y.~P.,  {White} S. D.~M.,  {B{\"o}rner} G.,
  {Cheng} F.~Z.,  2006, \mn@doi [\mnras] {10.1111/j.1365-2966.2006.10066.x},
  \href {https://ui.adsabs.harvard.edu/abs/2006MNRAS.368...21L} {368, 21}

\bibitem[\protect\citeauthoryear{{Lupton}, {Gunn}  \& {Szalay}}{{Lupton}
  et~al.}{1999}]{1999AJ....118.1406L}
{Lupton} R.~H.,  {Gunn} J.~E.,   {Szalay} A.~S.,  1999, \mn@doi [\aj]
  {10.1086/301004}, \href {http://adsabs.harvard.edu/abs/1999AJ....118.1406L}
  {118, 1406}

\bibitem[\protect\citeauthoryear{{Lupton}, {Gunn}, {Ivezi{\'c}}, {Knapp}  \&
  {Kent}}{{Lupton} et~al.}{2001}]{2001ASPC..238..269L}
{Lupton} R.,  {Gunn} J.~E.,  {Ivezi{\'c}} Z.,  {Knapp} G.~R.,   {Kent} S.,
  2001, in {Harnden} Jr. F.~R.,  {Primini} F.~A.,   {Payne} H.~E.,  eds,
  Astronomical Society of the Pacific Conference Series Vol. 238, Astronomical
  Data Analysis Software and Systems X. p.~269 (\mn@eprint {}
  {astro-ph/0101420})

\bibitem[\protect\citeauthoryear{{Lynds}}{{Lynds}}{1971}]{Field196}
{Lynds} R.,  1971, \mn@doi [\apj] {10.1086/180695}, \href
  {http://adsabs.harvard.edu/abs/1971ApJ...164L..73L} {164, L73}

\bibitem[\protect\citeauthoryear{{Madau}, {Ferrara}  \& {Rees}}{{Madau}
  et~al.}{2001}]{2001ApJ...555...92M}
{Madau} P.,  {Ferrara} A.,   {Rees} M.~J.,  2001, \mn@doi [\apj]
  {10.1086/321474}, \href
  {https://ui.adsabs.harvard.edu/abs/2001ApJ...555...92M} {555, 92}

\bibitem[\protect\citeauthoryear{{Martin}, {Chang}, {Matuszewski}, {Morrissey},
  {Rahman}, {Moore}  \& {Steidel}}{{Martin}
  et~al.}{2014a}]{2014ApJ...786..106M}
{Martin} D.~C.,  {Chang} D.,  {Matuszewski} M.,  {Morrissey} P.,  {Rahman} S.,
  {Moore} A.,   {Steidel} C.~C.,  2014a, \mn@doi [\apj]
  {10.1088/0004-637X/786/2/106}, \href
  {http://adsabs.harvard.edu/abs/2014ApJ...786..106M} {786, 106}

\bibitem[\protect\citeauthoryear{{Martin}, {Chang}, {Matuszewski}, {Morrissey},
  {Rahman}, {Moore}, {Steidel}  \& {Matsuda}}{{Martin}
  et~al.}{2014b}]{2014ApJ...786..107M}
{Martin} D.~C.,  {Chang} D.,  {Matuszewski} M.,  {Morrissey} P.,  {Rahman} S.,
  {Moore} A.,  {Steidel} C.~C.,   {Matsuda} Y.,  2014b, \mn@doi [\apj]
  {10.1088/0004-637X/786/2/107}, \href
  {http://adsabs.harvard.edu/abs/2014ApJ...786..107M} {786, 107}

\bibitem[\protect\citeauthoryear{{McDonald}, {Miralda-Escud{\'e}}, {Rauch},
  {Sargent}, {Barlow}, {Cen}  \& {Ostriker}}{{McDonald}
  et~al.}{2000}]{2000ApJ...543....1M}
{McDonald} P.,  {Miralda-Escud{\'e}} J.,  {Rauch} M.,  {Sargent} W. L.~W.,
  {Barlow} T.~A.,  {Cen} R.,   {Ostriker} J.~P.,  2000, \mn@doi [\apj]
  {10.1086/317079}, \href
  {https://ui.adsabs.harvard.edu/\#abs/2000ApJ...543....1M} {543, 1}

\bibitem[\protect\citeauthoryear{{McDonald} et~al.,}{{McDonald}
  et~al.}{2006}]{2006ApJS..163...80M}
{McDonald} P.,  et~al., 2006, \mn@doi [The Astrophysical Journal Supplement
  Series] {10.1086/444361}, \href
  {https://ui.adsabs.harvard.edu/\#abs/2006ApJS..163...80M} {163, 80}

\bibitem[\protect\citeauthoryear{{Miralda-Escud{\'e}}, {Cen}, {Ostriker}  \&
  {Rauch}}{{Miralda-Escud{\'e}} et~al.}{1996}]{1996ApJ...471..582M}
{Miralda-Escud{\'e}} J.,  {Cen} R.,  {Ostriker} J.~P.,   {Rauch} M.,  1996,
  \mn@doi [\apj] {10.1086/177992}, \href
  {http://adsabs.harvard.edu/abs/1996ApJ...471..582M} {471, 582}

\bibitem[\protect\citeauthoryear{{Moos} et~al.,}{{Moos}
  et~al.}{2000}]{2000ApJ...538L...1M}
{Moos} H.~W.,  et~al., 2000, \mn@doi [\apjl] {10.1086/312795}, \href
  {https://ui.adsabs.harvard.edu/abs/2000ApJ...538L...1M} {538, L1}

\bibitem[\protect\citeauthoryear{{Morris}, {Weymann}, {Savage}  \&
  {Gilliland}}{{Morris} et~al.}{1991}]{1991ApJ...377L..21M}
{Morris} S.~L.,  {Weymann} R.~J.,  {Savage} B.~D.,   {Gilliland} R.~L.,  1991,
  \mn@doi [\apj] {10.1086/186107}, \href
  {https://ui.adsabs.harvard.edu/\#abs/1991ApJ...377L..21M} {377, L21}

\bibitem[\protect\citeauthoryear{{Nicastro} et~al.,}{{Nicastro}
  et~al.}{2018}]{2018Natur.558..406N}
{Nicastro} F.,  et~al., 2018, \mn@doi [\nat] {10.1038/s41586-018-0204-1}, \href
  {https://ui.adsabs.harvard.edu/abs/2018Natur.558..406N} {558, 406}

\bibitem[\protect\citeauthoryear{{Ostriker}, {Bajtlik}  \& {Duncan}}{{Ostriker}
  et~al.}{1988}]{1988ApJ...327L..35O}
{Ostriker} J.~P.,  {Bajtlik} S.,   {Duncan} R.~C.,  1988, \mn@doi [\apjl]
  {10.1086/185135}, \href {http://adsabs.harvard.edu/abs/1988ApJ...327L..35O}
  {327, L35}

\bibitem[\protect\citeauthoryear{{Pallottini}, {Gallerani}  \&
  {Ferrara}}{{Pallottini} et~al.}{2014}]{2014MNRAS.444L.105P}
{Pallottini} A.,  {Gallerani} S.,   {Ferrara} A.,  2014, \mn@doi [\mnras]
  {10.1093/mnrasl/slu126}, \href
  {https://ui.adsabs.harvard.edu/abs/2014MNRAS.444L.105P} {444, L105}

\bibitem[\protect\citeauthoryear{Pan}{Pan}{2011}]{panthesis}
Pan D.,  2011, PhD thesis, Drexel University, PA 19104

\bibitem[\protect\citeauthoryear{{Pan}, {Vogeley}, {Hoyle}, {Choi}  \&
  {Park}}{{Pan} et~al.}{2012}]{2012MNRAS.421..926P}
{Pan} D.~C.,  {Vogeley} M.~S.,  {Hoyle} F.,  {Choi} Y.-Y.,   {Park} C.,  2012,
  \mn@doi [\mnras] {10.1111/j.1365-2966.2011.20197.x}, \href
  {https://ui.adsabs.harvard.edu/\#abs/2012MNRAS.421..926P} {421, 926}

\bibitem[\protect\citeauthoryear{{Penton}, {Stocke}  \& {Shull}}{{Penton}
  et~al.}{2002a}]{2002ApJ...565..720P}
{Penton} S.~V.,  {Stocke} J.~T.,   {Shull} J.~M.,  2002a, \mn@doi [\apj]
  {10.1086/324483}, \href {http://adsabs.harvard.edu/abs/2002ApJ...565..720P}
  {565, 720}

\bibitem[\protect\citeauthoryear{Penton, Stocke  \& Shull}{Penton
  et~al.}{2002b}]{Penton_2002}
Penton S.~V.,  Stocke J.~T.,   Shull J.~M.,  2002b, \mn@doi [The Astrophysical
  Journal] {10.1086/324483}, 565, 720

\bibitem[\protect\citeauthoryear{{Pierre}, {Shaver}  \& {Iovino}}{{Pierre}
  et~al.}{1988}]{1988A&A...197L...3P}
{Pierre} M.,  {Shaver} P.~A.,   {Iovino} A.,  1988, \aap, \href
  {http://adsabs.harvard.edu/abs/1988A%26A...197L...3P} {197, L3}

\bibitem[\protect\citeauthoryear{{Planck Collaboration} et~al.,}{{Planck
  Collaboration} et~al.}{2016}]{2016A&A...594A..22P}
{Planck Collaboration} et~al., 2016, \mn@doi [\aap]
  {10.1051/0004-6361/201525826}, \href
  {https://ui.adsabs.harvard.edu/\#abs/2016A&A...594A..22P} {594, A22}

\bibitem[\protect\citeauthoryear{{Planck Collaboration} et~al.,}{{Planck
  Collaboration} et~al.}{2018}]{2018arXiv180706209P}
{Planck Collaboration} et~al., 2018, arXiv e-prints, \href
  {https://ui.adsabs.harvard.edu/abs/2018arXiv180706209P} {p. arXiv:1807.06209}

\bibitem[\protect\citeauthoryear{{Prochaska} \& {Wolfe}}{{Prochaska} \&
  {Wolfe}}{2009}]{2009ApJ...696.1543P}
{Prochaska} J.~X.,  {Wolfe} A.~M.,  2009, \mn@doi [\apj]
  {10.1088/0004-637X/696/2/1543}, \href
  {https://ui.adsabs.harvard.edu/abs/2009ApJ...696.1543P} {696, 1543}

\bibitem[\protect\citeauthoryear{{Rauch} et~al.,}{{Rauch}
  et~al.}{1997}]{1997ApJ...489....7R}
{Rauch} M.,  et~al., 1997, \mn@doi [\apj] {10.1086/304765}, \href
  {https://ui.adsabs.harvard.edu/\#abs/1997ApJ...489....7R} {489, 7}

\bibitem[\protect\citeauthoryear{{Rodr{\'\i}guez Medrano}, {Paz}, {Stasyszyn}
  \& {Ruiz}}{{Rodr{\'\i}guez Medrano} et~al.}{2021}]{2021arXiv210913378R}
{Rodr{\'\i}guez Medrano} A.~M.,  {Paz} D.~J.,  {Stasyszyn} F.~A.,   {Ruiz}
  A.~N.,  2021, arXiv e-prints, \href
  {https://ui.adsabs.harvard.edu/abs/2021arXiv210913378R} {p. arXiv:2109.13378}

\bibitem[\protect\citeauthoryear{{Sargent}, {Young}, {Boksenberg}  \&
  {Tytler}}{{Sargent} et~al.}{1980}]{1980ApJS...42...41S}
{Sargent} W.~L.~W.,  {Young} P.~J.,  {Boksenberg} A.,   {Tytler} D.,  1980,
  \mn@doi [\apjs] {10.1086/190644}, \href
  {http://adsabs.harvard.edu/abs/1980ApJS...42...41S} {42, 41}

\bibitem[\protect\citeauthoryear{{Shull}, {Smith}  \& {Danforth}}{{Shull}
  et~al.}{2012}]{2012ApJ...759...23S}
{Shull} J.~M.,  {Smith} B.~D.,   {Danforth} C.~W.,  2012, \mn@doi [\apj]
  {10.1088/0004-637X/759/1/23}, \href
  {http://adsabs.harvard.edu/abs/2012ApJ...759...23S} {759, 23}

\bibitem[\protect\citeauthoryear{{Spitzer}, {Drake}, {Jenkins}, {Morton},
  {Rogerson}  \& {York}}{{Spitzer} et~al.}{1973}]{1973ApJ...181L.116S}
{Spitzer} L.,  {Drake} J.~F.,  {Jenkins} E.~B.,  {Morton} D.~C.,  {Rogerson}
  J.~B.,   {York} D.~G.,  1973, \mn@doi [\apjl] {10.1086/181197}, \href
  {https://ui.adsabs.harvard.edu/abs/1973ApJ...181L.116S} {181, L116}

\bibitem[\protect\citeauthoryear{{Steidel}, {Bogosavljevi{\'c}}, {Shapley},
  {Kollmeier}, {Reddy}, {Erb}  \& {Pettini}}{{Steidel}
  et~al.}{2011}]{2011ApJ...736..160S}
{Steidel} C.~C.,  {Bogosavljevi{\'c}} M.,  {Shapley} A.~E.,  {Kollmeier} J.~A.,
   {Reddy} N.~A.,  {Erb} D.~K.,   {Pettini} M.,  2011, \mn@doi [\apj]
  {10.1088/0004-637X/736/2/160}, \href
  {http://adsabs.harvard.edu/abs/2011ApJ...736..160S} {736, 160}

\bibitem[\protect\citeauthoryear{{Stocke}, {Penton}, {Danforth}, {Shull},
  {Tumlinson}  \& {McLin}}{{Stocke} et~al.}{2006}]{2006ApJ...641..217S}
{Stocke} J.~T.,  {Penton} S.~V.,  {Danforth} C.~W.,  {Shull} J.~M.,
  {Tumlinson} J.,   {McLin} K.~M.,  2006, \mn@doi [\apj] {10.1086/500386},
  \href {https://ui.adsabs.harvard.edu/\#abs/2006ApJ...641..217S} {641, 217}

\bibitem[\protect\citeauthoryear{{Stocke}, {Keeney}, {Danforth}, {Oppenheimer},
  {Pratt}, {Berlind}, {Impey}  \& {Jannuzi}}{{Stocke}
  et~al.}{2019}]{2019ApJS..240...15S}
{Stocke} J.~T.,  {Keeney} B.~A.,  {Danforth} C.~W.,  {Oppenheimer} B.~D.,
  {Pratt} C.~T.,  {Berlind} A.~A.,  {Impey} C.,   {Jannuzi} B.,  2019, \mn@doi
  [\apjs] {10.3847/1538-4365/aaf73d}, \href
  {https://ui.adsabs.harvard.edu/abs/2019ApJS..240...15S} {240, 15}

\bibitem[\protect\citeauthoryear{{Strauss} et~al.,}{{Strauss}
  et~al.}{2002}]{2002AJ....124.1810S}
{Strauss} M.~A.,  et~al., 2002, \mn@doi [\aj] {10.1086/342343}, \href
  {http://adsabs.harvard.edu/abs/2002AJ....124.1810S} {124, 1810}

\bibitem[\protect\citeauthoryear{{Tanimura}, {Aghanim}, {Douspis}, {Beelen}  \&
  {Bonjean}}{{Tanimura} et~al.}{2018}]{2018arXiv180504555T}
{Tanimura} H.,  {Aghanim} N.,  {Douspis} M.,  {Beelen} A.,   {Bonjean} V.,
  2018, arXiv e-prints, \href
  {https://ui.adsabs.harvard.edu/\#abs/2018arXiv180504555T} {p.
  arXiv:1805.04555}

\bibitem[\protect\citeauthoryear{{Tejos}, {Morris}, {Crighton}, {Theuns},
  {Altay}  \& {Finn}}{{Tejos} et~al.}{2012}]{2012MNRAS.425..245T}
{Tejos} N.,  {Morris} S.~L.,  {Crighton} N.~H.~M.,  {Theuns} T.,  {Altay} G.,
  {Finn} C.~W.,  2012, \mn@doi [\mnras] {10.1111/j.1365-2966.2012.21448.x},
  \href {http://adsabs.harvard.edu/abs/2012MNRAS.425..245T} {425, 245}

\bibitem[\protect\citeauthoryear{{Theuns}, {Leonard}  \& {Efstathiou}}{{Theuns}
  et~al.}{1998}]{1998MNRAS.297L..49T}
{Theuns} T.,  {Leonard} A.,   {Efstathiou} G.,  1998, \mn@doi [\mnras]
  {10.1046/j.1365-8711.1998.01740.x}, \href
  {https://ui.adsabs.harvard.edu/\#abs/1998MNRAS.297L..49T} {297, L49}

\bibitem[\protect\citeauthoryear{{Tonnesen}, {Smith}, {Kollmeier}  \&
  {Cen}}{{Tonnesen} et~al.}{2017}]{2017ApJ...845...47T}
{Tonnesen} S.,  {Smith} B.~D.,  {Kollmeier} J.~A.,   {Cen} R.,  2017, \mn@doi
  [\apj] {10.3847/1538-4357/aa7fb8}, \href
  {https://ui.adsabs.harvard.edu/abs/2017ApJ...845...47T} {845, 47}

\bibitem[\protect\citeauthoryear{{Viel}, {Haehnelt}  \& {Springel}}{{Viel}
  et~al.}{2004}]{2004MNRAS.354..684V}
{Viel} M.,  {Haehnelt} M.~G.,   {Springel} V.,  2004, \mn@doi [\mnras]
  {10.1111/j.1365-2966.2004.08224.x}, \href
  {https://ui.adsabs.harvard.edu/\#abs/2004MNRAS.354..684V} {354, 684}

\bibitem[\protect\citeauthoryear{{Viel}, {Becker}, {Bolton}, {Haehnelt},
  {Rauch}  \& {Sargent}}{{Viel} et~al.}{2008}]{2008PhRvL.100d1304V}
{Viel} M.,  {Becker} G.~D.,  {Bolton} J.~S.,  {Haehnelt} M.~G.,  {Rauch} M.,
  {Sargent} W. L.~W.,  2008, \mn@doi [\prl] {10.1103/PhysRevLett.100.041304},
  \href {https://ui.adsabs.harvard.edu/\#abs/2008PhRvL.100d1304V} {100, 041304}

\bibitem[\protect\citeauthoryear{{Weinberg}, {Miralda-Escud{\'e}}, {Hernquist}
  \& {Katz}}{{Weinberg} et~al.}{1997}]{1997ApJ...490..564W}
{Weinberg} D.~H.,  {Miralda-Escud{\'e}} J.,  {Hernquist} L.,   {Katz} N.,
  1997, \mn@doi [\apj] {10.1086/304893}, \href
  {https://ui.adsabs.harvard.edu/\#abs/1997ApJ...490..564W} {490, 564}

\bibitem[\protect\citeauthoryear{{Weinberg}, {Katz}  \& {Hernquist}}{{Weinberg}
  et~al.}{1998}]{1998ASPC..148...21W}
{Weinberg} D.~H.,  {Katz} N.,   {Hernquist} L.,  1998, in {Woodward} C.~E.,
  {Shull} J.~M.,   {Thronson} Jr. H.~A.,  eds,  Astronomical Society of the
  Pacific Conference Series Vol. 148, Origins. p.~21 (\mn@eprint {}
  {astro-ph/9708213})

\bibitem[\protect\citeauthoryear{{Wilde} et~al.,}{{Wilde}
  et~al.}{2021}]{2021ApJ...912....9W}
{Wilde} M.~C.,  et~al., 2021, \mn@doi [\apj] {10.3847/1538-4357/abea14}, \href
  {https://ui.adsabs.harvard.edu/abs/2021ApJ...912....9W} {912, 9}

\bibitem[\protect\citeauthoryear{{Zhang}, {Anninos}  \& {Norman}}{{Zhang}
  et~al.}{1995}]{1995ApJ...453L..57Z}
{Zhang} Y.,  {Anninos} P.,   {Norman} M.~L.,  1995, \mn@doi [\apjl]
  {10.1086/309752}, \href {http://adsabs.harvard.edu/abs/1995ApJ...453L..57Z}
  {453, L57}

\bibitem[\protect\citeauthoryear{{de Graaff}, {Cai}, {Heymans}  \&
  {Peacock}}{{de Graaff} et~al.}{2017}]{2017arXiv170910378D}
{de Graaff} A.,  {Cai} Y.-C.,  {Heymans} C.,   {Peacock} J.~A.,  2017, arXiv
  e-prints, \href {https://ui.adsabs.harvard.edu/\#abs/2017arXiv170910378D} {p.
  arXiv:1709.10378}

\makeatother
\end{thebibliography}

\bsp
\label{lastpage}
\end{document}